\documentclass[aip,reprint]{revtex4-1}
\usepackage{amsthm, amsmath, amssymb, dsfont, graphics, graphicx}

\newtheorem{definition}{Definition}[section]

\DeclareMathOperator{\E}{E}
\DeclareMathOperator{\Law}{Law}
\DeclareMathOperator{\Var}{Var}

\DeclareMathOperator{\Prob}{P}
\DeclareMathOperator{\Unif}{Unif}
\DeclareMathOperator{\anc}{anc}

\begin{document}


\title[Practical rare event sampling]{Practical rare event sampling for extreme mesoscale weather} 



\author{Robert J. Webber}
\email{rw2515@nyu.edu}
\affiliation{Courant Institute of Mathematical Sciences, New York University, New York, NY 10012}
\author{David A. Plotkin}
\affiliation{Department of the Geophysical Sciences, University of Chicago, Chicago, IL 60637}
\author{Morgan E O'Neill}
\affiliation{Department of Earth System Science, Stanford University, Stanford, CA 94305}
\author{Dorian S. Abbot}
\affiliation{Department of the Geophysical Sciences, University of Chicago, Chicago, IL 60637}
\author{Jonathan Weare}
\affiliation{Courant Institute of Mathematical Sciences, New York University, New York, NY 10012}


\date{\today}

\begin{abstract}
Extreme mesoscale weather, including tropical cyclones, squall lines, and floods, 
can be enormously damaging and yet
challenging to simulate;
hence, there is a pressing need for more efficient simulation strategies.
Here we present a new rare event sampling algorithm called 
Quantile Diffusion Monte Carlo (Quantile DMC).
Quantile DMC is a simple-to-use algorithm that can sample extreme tail behavior
for a wide class of processes.
We demonstrate the advantages of Quantile DMC compared to other sampling methods
and discuss practical aspects of implementing Quantile DMC.
To test the feasibility of Quantile DMC for extreme mesoscale weather,
we sample extremely intense 
realizations of two historical tropical cyclones, 2010 Hurricane Earl
and 2015 Hurricane Joaquin.
Our results demonstrate Quantile DMC's potential 
to provide low-variance extreme weather statistics
while highlighting the work that is necessary for Quantile DMC 
to attain greater efficiency in future applications.
\end{abstract}


\maketitle 


\begin{quotation}
When rare events are studied using simulation,
it can take a long time to gather sufficient data through direct sampling.
As an alternative to direct sampling,
specialized rare event sampling algorithms provide data more quickly,
thus reducing computational costs.
Here, we present a new rare event sampling method, Quantile DMC, 
that is simple to use.
Quantile DMC performs extremely well
on a one-dimensional test case,
accurately estimating rare event probabilities
with less than one thousandth the computational cost of direct sampling.
Quantile DMC could potentially be of use
in complex rare event simulations,
for example, simulating the frequency of
tropical cyclones, mesoscale convective systems, or floods
under different climate conditions.
When we apply
Quantile DMC to simulate intense tropical cyclones,
we obtain promising results:
storms at high intensities are more reliably simulated
using Quantile DMC compared to direct sampling.
\end{quotation}

\section{Introduction}{\label{sec:Introduction}}

A common strategy for estimating rare event probabilities is \emph{direct sampling} \cite{meehl2004more, bender2010modeled, lin2016grey}.
The direct sampling approach is to repeatedly simulate data from a model
and then calculate the frequency of a rare event over all the simulated data.
This approach can be effective in some contexts
but can also be computationally expensive.
For the rarest probabilities,
an exorbitant amount of computational effort
might be required before the event occurs even once in the simulations.
Responding to these concerns, researchers as early as the 1950s \cite{kahn1951estimation, rosenbluth1955monte}
developed specialized rare event sampling algorithms 
to improve computational efficiency.

Today, a diverse community of scientists uses rare event sampling 
and analysis tools to study processes that take place infrequently, 
are too complex to be described analytically, and can be simulated on a computer.
For example, the following extraordinary events have all been simulated
using rare event sampling:
a high-energy particle penetrating a nuclear shield \cite{dubi1982geometrical},
a life-sustaining protein-protein reaction \cite{huber1996weighted},
and an extreme loss of portfolio value \cite{glasserman2000variance}.

A burgeoning field of research explicitly links rare event simulation
and analysis tools with geophysical applications 
\cite{hoffman2006response, weare2009particle, vanden2013data, 
ragone2017computation, dematteis2018rogue, plotkin2018maximizing}.
In a recent paper,
Ragone, Wouters and Bouchet \cite{ragone2017computation}
showed that rare event sampling methods can be used
to study the probability of extreme weather occurring.
They sampled intense 90-day heat waves over Western Europe
at a fraction of the computational expense of direct sampling.
Their simulations led to the surprising insight that
extreme heat waves over Western Europe are associated
with a stationary wavenumber 3
anomaly in the jet stream.

The work of Ragone and coauthors concerns
\emph{synoptic scale} weather,
weather that occurs on a length scale of 1000km or greater
\cite{meehl2004more}.
A more challenging question is how to apply rare event sampling 
techniques to \emph{mesoscale} weather,
which occurs on a smaller length scale of 10-1000km.
Extreme mesoscale weather, including tropical cyclones and floods,
accounts for many of the world's most destructive natural disasters 
\cite{pielke2000precipitation, pielke2008normalized}.
Yet simulations of mesoscale weather can demand enormous computational resources
due to the need for high spatial resolution \cite{davis2008prediction, hirabayashi2008global}.
Mesoscale weather simulation presents a unique challenge for rare event sampling
and analysis,
where the need for efficient simulation strategies is great and yet
sample size is highly limited due to computational expense.

In \citet{plotkin2018maximizing},
we present a rare event analysis strategy
for potential use in extreme mesoscale weather simulations.
Using a computationally efficient algorithm,
we identify maximum likelihood perturbations that
lead to the occurrence of an intense tropical cyclone
in a high-resolution weather model.
In particular, we identify key changes in wind, temperature, and relative humidity fields
that help explain the rapid intensification process in modeled tropical cyclones.

In contrast to \citet{plotkin2018maximizing},
which analyzes the single most likely path toward
extreme mesoscale weather,
the present paper analyzes \emph{statistics} of
extreme mesoscale weather.
Accurate estimation of statistics
can require sampling numerous possible paths.
To achieve this goal, therefore, 
it is appropriate to use a rare event sampling algorithm.

Here we present a new rare event sampling algorithm
called Quantile Diffusion Monte Carlo (Quantile DMC)
that is suited
for complex real-world applications
such as extreme weather simulations.
The algorithm is simple to implement,
yet suitable for a large class of nonlinear processes.
When we apply Quantile DMC
to a simple example,
the algorithm is more efficient than direct sampling by a factor of more than a thousand.

Quantile DMC is a ``splitting" algorithm,
an algorithm in which some simulations are split into multiple replicas
to promote progress toward the rare event of interest
and other simulations are ``killed''.
A key advantage of splitting algorithms is that they are practical to implement.
Our simulations with Quantile DMC are simple to code
and require the same computational cost as direct sampling from the dynamical model.
In contrast, alternative rare event sampling approaches can be more challenging to implement,
because they require modifying the underlying dynamical model
or frequently starting and stopping the dynamics.
For mesoscale weather models like the one simulated in the current paper, 
these manipulations would require extensive code development
or substantial added computational cost.

Quantile DMC is inspired by a previous splitting algorithm called
Diffusion Monte Carlo 
\cite{kahn1951estimation, rosenbluth1955monte, kalos1962monte, grimm1971monte, giardina2006direct, hairer2014improved},
but it incorporates two new features.
First, in Diffusion Monte Carlo,
splitting is typically uniform in time,
but in Quantile DMC the intensity and frequency of splitting
increase over time, improving efficiency.
Second, Quantile DMC
adaptively makes use of data from simulations,
so that the algorithm requires
less tuning compared to DMC.

We envision that Quantile DMC could be used to
study the frequency of extreme mesoscale weather
under different climate conditions.
While a full application of rare event sampling techniques
to study extreme weather
is beyond the scope of the current paper,
we test the feasibility of our approach using a high-resolution
tropical cyclone model.
In simulations,
we use Quantile DMC to study the upper tail of the intensity distribution for 
numerical simulations of two historical tropical cyclones: 2010 Hurricane Earl
and 2015 Hurricane Joaquin.
Using an ensemble of $N = 100$ simulations, 
Quantile DMC produces more than seven times as many high-intensity Category 5 realizations
for both of the storms
compared to direct sampling.
Moreover, the variance of important rare event statistics is improved
by a factor of two to ten.
Building on this success, we anticipate
that we can improve the performance of this method in the future.

This paper is organized into two major sections.
Section \ref{sec:Math} presents various approaches
to estimating rare event probabilities:
direct sampling, Diffusion Monte Carlo,
and the new method Quantile DMC.
Section \ref{sec:Extreme} examines the potential role of rare event sampling
in extreme weather and presents tropical cyclone simulations.


\section{Estimating rare event probabilities}{\label{sec:Math}}

In this section, we introduce several approaches to estimating rare event probabilities
for potential use in climate and weather applications.
First, we discuss the well-known method of direct sampling.
Then, we describe a rare event sampling algorithm, called Diffusion Monte Carlo (DMC).
We illustrate the advantages and disadvantages of DMC on a simple example.
Then, we introduce Quantile DMC and explain why it gives more
robust performance compared to standard DMC.
Lastly, we discuss the implementation of Quantile DMC in practical settings.

\subsection{Direct sampling}


While direct sampling is a very useful tool for studying the typical behavior
of complex or high-dimensional systems, 
it is not an ideal approach for investigating unlikely or
infrequent phenomena.
In particular, as we will demonstrate, direct sampling can give very high error
when estimating statistics of rare events.

Direct sampling uses a straightforward approach to estimate probabilities. 
We assume $X$ is a random process, 
$A$ is an important event
(e.g., the occurrence of an intense tropical cyclone),
and we can draw independent samples of $X$, labeled
$\xi^{\left(1\right)}, \xi^{\left(2\right)}, \xi^{\left(3\right)}, \ldots, \xi^{\left(N\right)}$.
To estimate the probability
$p = \Prob\left\{X \in A\right\}$,
direct sampling uses
\begin{equation}
\label{directestimates}
\hat{p} =
\frac{1}{N} \sum_{j=1}^N \mathds{1}\left\{\xi^{\left(j\right)} \in A\right\}
\end{equation}
This estimator is called the \emph{sample average}.

To assess the error in direct sampling, we calculate the mean and variance of $\hat{p}$:
\begin{equation}{\label{error}}
\begin{cases}
\E \left[\hat{p}\right] = p \\
\Var \left[\hat{p}\right] = \frac{1}{N} p\left(1-p\right)
\end{cases}
\end{equation}

On the surface, the variance of the sample average $\hat{p}$ would appear to be quite good.
In particular, the variance depends only on $p$ and not
on the process $X$.
The process $X$ can have millions of dimensions
or even infinite dimensions
and still the variance of $\hat{p}$ converges to zero at a
$1 \slash N$ rate as $N \rightarrow \infty$.

Surprisingly then, direct sampling estimates $\hat{p}$
can have unacceptably high error in rare event calculations.
For example, suppose $p = 0.01$ is the probability of a rare event $A$,
and $N = 100$ is the sample size of simulations.
Then,
$\hat{p}$ may take the value $\hat{p} = 0$ with probability $0.37$,
the value $\hat{p} = 0.01$ with probability $0.37$,
the value $\hat{p} = 0.02$ with probability $0.18$,
the value $\hat{p} = 0.03$ with probability $0.06$,
and higher values with probability $0.02$.
The error in $\hat{p}$ is overwhelming.
If the estimate $\hat{p}$ is used for risk analysis,
then the error in $\hat{p}$ might have harmful practical consequences.

How can direct sampling produce estimates that are 
simultaneously so good and so bad?
In most applications,
what is important is not absolute error $\hat{p} - p$,
but rather \emph{relative error} $\left(\hat{p} - p\right) \slash p$.
To assess the relative error in direct sampling, we calculate the mean and variance of $\hat{p} \slash p$:
\begin{equation}{\label{newerror}}
\begin{cases}
\E\left[\hat{p} \slash p\right] = 1 \\
\Var\left[\hat{p} \slash p\right] = \frac{1-p}{Np}
\end{cases}
\end{equation}
To estimate a probability $p$ with even one digit of precision, 
$\Var\left[\hat{p} \slash p\right]$ must be many times smaller than $1$, 
and this requires a sample size $N$ that is many
times larger than $1 \slash p$.
When models are expensive to run and probabilities $p$ are small, obtaining these
large sample sizes is not a practical option.

Rare event sampling methods can help address
the deficiency of direct sampling in estimating rare probabilities.
For example, in carefully designed
rare event splitting algorithms similar to Diffusion Monte Carlo and Quantile DMC
it can suffice to increase sample size $N$ as slowly as $N \sim \log\left(1 \slash p\right)$
as $p \rightarrow 0$
to achieve fixed relative error \cite{cerou2006genetic}.
This is an exponential improvement compared to direct sampling,
where it is necessary to increase sample size $N$ at a rate of $N \sim 1 \slash p$
to achieve fixed relative error.
The exponential improvement due to rare event sampling methods
can make possible very precise
calculations of rare event probabilities even with a limited ensemble size $N$.


\subsection{Diffusion Monte Carlo}

Diffusion Monte Carlo (DMC) is a sampling algorithm that causes simulations to explore regions
of state space that would rarely be accessed under typical conditions.
The earliest antecedents of DMC were splitting algorithms
invented in the 1950s \cite{kahn1951estimation, rosenbluth1955monte}.
In the 1960s, DMC was popularized in the quantum chemistry community
where researchers used DMC to obtain information about the ground state
energy of the Schr\"{o}dinger equation for chemical systems \cite{kalos1962monte, grimm1971monte}.
In the 2000s, the tools of DMC were increasingly applied to rare event sampling,
and the algorithm became known in some circles as ``genealogical particle analysis''
\cite{del2005genealogical, wouters2016rare}.
Recently, DMC has been the subject of a series of mathematical analyses,
which describe the convergence and asymptotic behavior 
of DMC as the ensemble size $N$ approaches infinity \cite{del2004feynman, webber2019resampling}.
Hairer and Weare \cite{hairer2014improved} provide a more detailed
history of DMC.


For a simple example of DMC, 
assume $\left(X_t\right)_{t \geq 0}$ is a Markov process in $\mathbb{R}^d$, 
and $\xi_t^{\left(1\right)}, \xi_t^{\left(2\right)}, \ldots, \xi_t^{\left(N\right)}$
are simulations of $X_t$, which are called ``particles''.
To estimate the probability of a rare, important event $A$,
the DMC algorithm iterates the following steps:
\begin{enumerate}
\item
Evolve particles $\left(\xi_t^{\left(i\right)}\right)_{1 \leq i \leq N}$ forward from time $t$ to a later time $t^{\prime}$.
\item
Using a consistent set of rules,
randomly ``split'' particles $\xi_{t^{\prime}}^{\left(i\right)}$ that have moved much closer to $A$
and randomly ``kill'' particles $\xi_{t^{\prime}}^{\left(i\right)}$ that have moved much farther from $A$,
making sure that the total number of particles $N$ remains unchanged.
\end{enumerate}
DMC uses splitting and killing to cause
a greater number of particles to reach the rare event state $A$,
compared to direct sampling.

In greater generality, the DMC algorithm is guided by one-dimensional coordinate
$\theta\colon \mathbb{R}^d \rightarrow \mathbb{R}$
that is high in some regions of the state space $\mathbb{R}^d$
and low in other regions of the state space.
Where $\theta$ is high, DMC exhibits a greater propensity toward splitting.
Where $\theta$ is low, DMC exhibits a greater propensity toward killing.
The coordinate $\theta$ is often known
as an \emph{order parameter} or \emph{reaction coordinate}.
The particular choice of reaction coordinate can be crucial to the efficiency of DMC
for computing rare event statistics.

\begin{figure}[!htbp]
\includegraphics[width=8cm, trim = {0 0 0 .1cm}, clip]{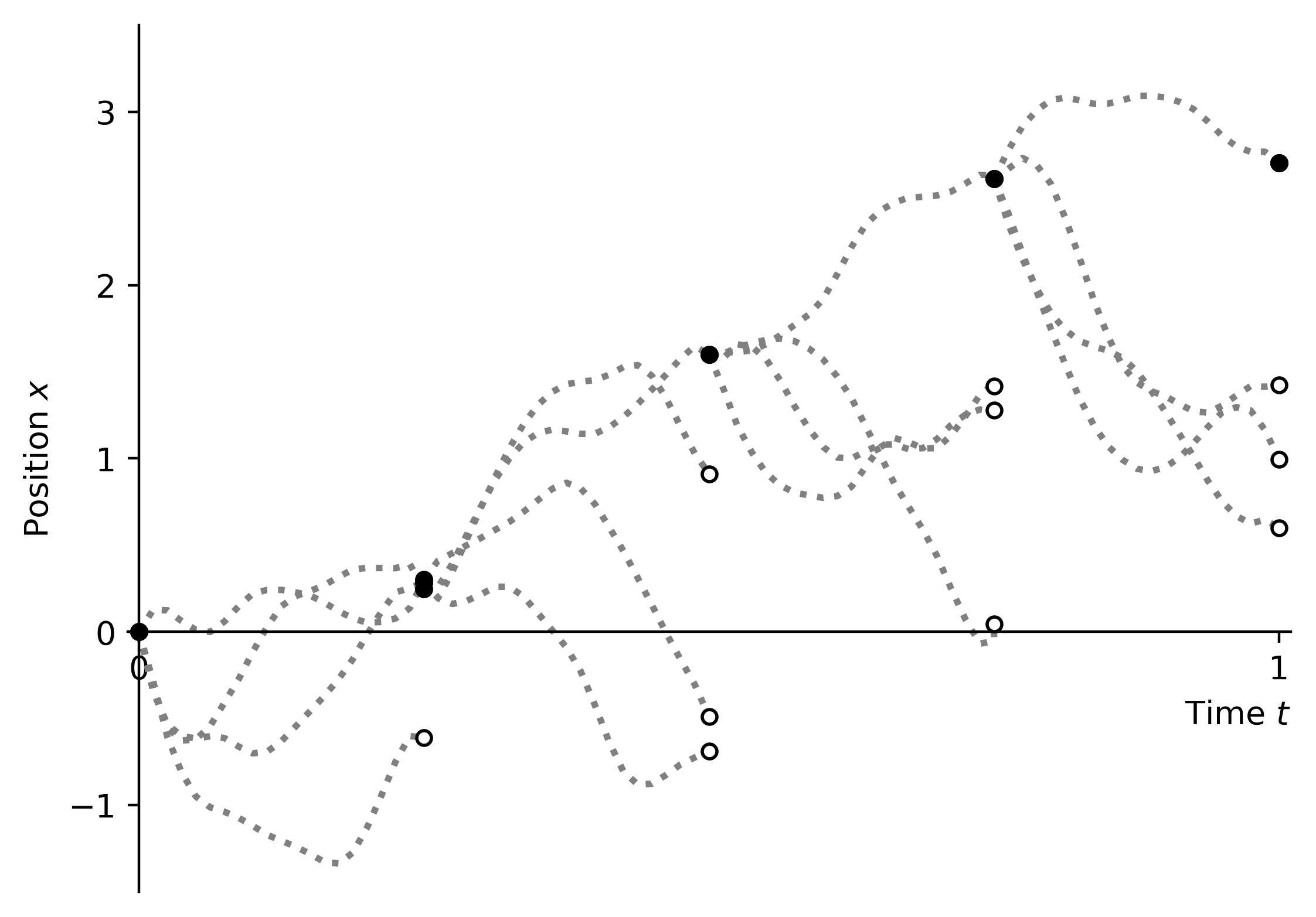}
\caption{Illustration of Diffusion Monte Carlo.
At fixed times $t$
some simulations are killed (white circles).
Simulations that are not killed (black circles)
are possibly replicated, and
all simulations are run forward in time.
Splitting and killing create a net flux,
driving simulations toward high values of $x$.}
\label{figure1}
\end{figure}

A basic schematic of DMC is given in Figure \ref{figure1}.
In this schematic, the reaction coordinate is the position $\theta\left(x\right) = x$.
Therefore, splitting and killing of simulations drives the process toward high values of $x$.

To implement DMC on a computer,
a sequence of actions are required.
The user defines a series of resampling times
$0 = t_0 < t_1 < t_2 < \cdots$.
For each resampling time $t_k$,
the user specifies $V_k\left(x\right)$,
a splitting function that increases with the reaction coordinate $\theta\left(x\right)$.
DMC begins with an initialization step and then
iterates over reweighting, resampling, and mutation steps
according to the following definition:

\begin{definition}[Diffusion Monte Carlo]{\label{def:DMC}}
~
\begin{enumerate}
\item Initialization: Independently sample initial particles
$\xi_{0}^{\left(i\right)}\sim\Law\left(X_{0}\right)$ for
$1 \leq i \leq N$
\item For $k=0,1,2,\ldots$,
\begin{enumerate}
\item Reweighting: If $k = 0$, define initial weights
\begin{equation}
w_{0}^{\left(i\right)} = \exp\left\{V_0\left(\xi_0^{\left(i\right)}\right)\right\}
\end{equation}
If $k > 0$, define weights
\begin{flalign}
&& w_k^{\left(i\right)}
= \overline{w}_{k-1}
\exp\left\{V_k\left(\xi_k^{\left(i\right)}\right) - 
V_{k-1}\left(\hat{\xi}_{k-1}^{\left(i\right)}\right)\right\}
\end{flalign}
Define the average weight $\overline{w}_k = \frac{1}{N} \sum_{i=1}^N w_k^{\left(i\right)}$.
\item Resampling:
By splitting and killing particles $\left(\xi_k^{\left(i\right)}\right)_{1 \leq i \leq N}$,
create an ensemble of updated particles $\left(\hat{\xi}_k^{\left(j\right)}\right)_{1 \leq j \leq N}$
consisting of $N_k^{\left(i\right)}$ copies of each particle $\xi_k^{\left(i\right)}$.
The numbers $N_k^{\left(i\right)}$ are randomly chosen to satisfy
\begin{equation}
\label{condition}
\begin{cases}
\sum_{i=1}^N N_k^{\left(i\right)} = N \\
\E \left[N_k^{\left(i\right)}\right] = w_k^{\left(i\right)} \slash \overline{w}_k
\end{cases}
\end{equation}
\item Mutation: Independently sample
$\xi_{k+1}^{\left(i\right)}\sim \Law \left(X_{t_{k+1}}|X_{t_k}=\hat{\xi}_k^{\left(i\right)}\right)$
for $1\leq i\leq N$.
\end{enumerate}
\item Estimation: To approximate $\E \left[f\left(X_{t_k}\right)\right] $, 
DMC uses the estimate
\begin{equation}
\E \left[f\left(X_{t_k}\right)\right]
= \frac{\overline{w}_{k-1}}{N} \sum_{i=1}^N \frac{f\left(\xi_k^{\left(i\right)}\right)}
{\exp\left\{V_{k-1}\left(\hat{\xi}_{k-1}^{\left(j\right)}\right)\right\}}
\end{equation}
\end{enumerate}
\end{definition}

The intialization and mutation steps in DMC are straightforward,
but the reweighting and resampling steps require further elaboration.
In the reweighting step, splitting/killing weights $\left(w_k^{\left(i\right)}\right)_{1 \leq i \leq N}$
are defined by means of the splitting functions $V_k$.
The simplest example of a splitting function is $V_k\left(x\right) = C \theta\left(x\right)$, 
where $C > 0$
is a splitting parameter that controls how many times a single particle can be split to create new copies.

In the resampling step, particles are split and killed
according to the weights $\left(w_k^{\left(i\right)}\right)_{1 \leq i \leq N}$.
Random numbers $N_k^{\left(i\right)}$
indicate how many times each particle $\xi_k^{\left(i\right)}$ is copied.
The random numbers $N_k^{\left(i\right)}$ have expectation
\begin{equation}
\E \left[N_k^{\left(i\right)}\right] = w_k^{\left(i\right)} \slash \overline{w}_k
\end{equation}
In this formula, $w_k^{\left(i\right)}$ is divided by $\overline{w}_k$ is to ensure that
the total number of particles satisfies
$\E \left[\sum_{i=1}^N N_k^{\left(i\right)}\right] = N$.
To define the particular distribution for the random numbers $\left(N_k^{\left(i\right)}\right)_{1 \leq i \leq N}$,
we use a low-variance resampling scheme called \emph{sorted stratified resampling} \cite{webber2019resampling, kitagawa1996monte},
the details of which we describe in the appendix.
While a good choice of resampling scheme can slighly reduce DMC error,
other factors determine a greater share of the error in DMC estimates.
The splitting functions $V_k$ are the most 
important parameters for determining the dynamics of DMC.


The validity of DMC estimates 
is supported by mathematical analyses \cite{del2004feynman, webber2019resampling}.
DMC estimates are unbiased
and converge as the number of particles $N$ tends to infinity
under mild integrability conditions.
These theoretical results are very general,
holding true for systems with arbitrarily high dimension $d$.
Any quantity that can be estimated by direct sampling
can also be estimated by DMC.
Estimates can include functions that depend on the entire path
from time $0$ until a later time $t_k$.


Analysis of DMC supports
the conclusion that DMC oversamples regions where the reaction coordinate $\theta$ is large
and undersamples regions where $\theta$ is small.
In particular, the distribution of particles 
$\frac{1}{N} \sum_{j=1}^{N} \delta\left(\hat{\xi}_k^{\left(j\right)}\right)$
converges weakly as $N \rightarrow \infty$ to the distribution of $X_{t_k}$ weighted by a likelihood ratio of
\begin{equation}{\label{likelihood}}
L_k\left(x\right) = \frac{\exp\left\{V_k\left(x\right)\right\}}
{\E\left[\exp\left\{V_k\left(X_{t_k}\right)\right\}\right]}
\end{equation}
Since $V_k\left(x\right)$ increases with the reaction coordinate $\theta\left(x\right)$,
more particles occupy regions where $\theta$ is high,
compared to direct sampling.

In summary, we have defined Diffusion Monte Carlo and presented two key facts.
First, DMC provides unbiased, convergent estimates.
Second, at each resampling step $t_k$,
DMC moves particles to regions where the reaction coordinate $\theta$ is large.


\subsection{Strengths and weaknesses of DMC: the Ornstein-Uhlenbeck example}{\label{sub:example}}

\begin{figure*}[t]
\includegraphics[width=16cm, trim = {0 0 0 .1cm}, clip]{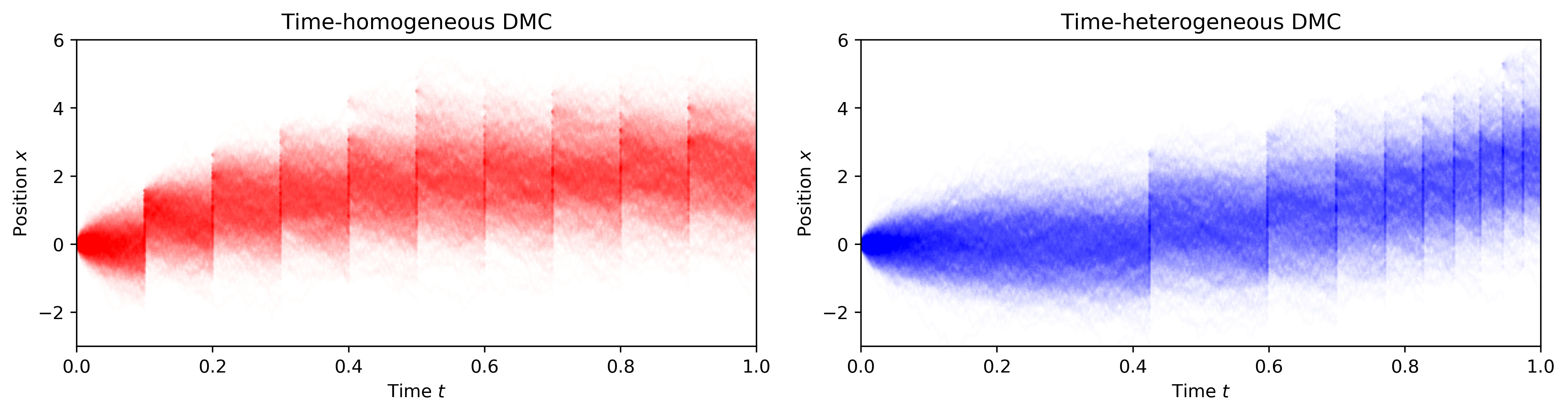}
\caption{In time-homogeneous DMC (left)
particles reach high positions $x$
more quickly than in time-heterogeneous DMC (right).
Both methods are implemented with $N = 1000$ particles,
tilting constant $C = 2.5$,
and dynamics
$\mathop{dX_t} = -2X_t \mathop{dt} + 2\mathop{dW_t}$.}
\label{figure2}
\end{figure*}

The Ornstein-Uhlenbeck (OU) process is a one-dimensional, linear process which we use to illustrate
the strengths and weaknesses of DMC.
On the one hand, DMC can effectively sample rare extreme deviations of the OU process.
On the other hand, when sampling transformations of the OU process,
DMC can require delicate tuning,
which limits the practical effectiveness of the algorithm.

The OU process evolves under the dynamics
\begin{equation}
\mathop{dX_t} = -\alpha X_t \mathop{dt} + \sqrt{2 \alpha} \mathop{dW_t}
\end{equation}
where $\alpha > 0$ is a constant.
We present three properties of the OU process for later reference:

\begin{enumerate}
\item
From any starting distribution, the OU process converges geometrically to an equilibrium distribution of
$N\left(0, 1\right)$.
Therefore, it is a rare event for the OU process to reach a position much larger than the standard deviation of $1$
at a large time $T$.
\item
From any starting position $X_0 = x$, the mean of the OU process converges geometrically to $0$ at a timescale of $1 \slash \alpha$, that is, $\E \left[X_t\right] = x e^{-\alpha t}$.
\item
From any starting position $X_0 = x$, the variance of the OU process converges geometrically to $1$ at a timescale of $1 \slash \left(2 \alpha\right)$, that is,
$\Var \left[X_t\right] = 1 - e^{-2 \alpha t}$.
\end{enumerate}

We consider using DMC to estimate the probability that 
the OU process starting from $X_0 = 0$ exhibits a rare extreme deviation $X_1 \geq U$.
To set up the DMC algorithm, a natural choice of reaction coordinate is position $\theta\left(x\right) = x$.
We must also choose
a series of resampling times $0 < t_1 < \cdots < t_{K-1} < t_K = 1$
and splitting functions $V_k$.
The simplest choice is to set $t_k = k \slash K$ for $k = 1, 2, \ldots, K$
and $V_k\left(x\right) = C \theta\left(x\right)$, where $C > 0$ is a positive number.
We refer to this strategy as ``time-homogeneous'' DMC,
since the splitting intensity and splitting frequency are uniform in time.


As an alternative to time-homogeneous DMC,
we also consider a ``time-heterogeneous'' resampling strategy.
Since the OU process loses its memory exponentially quickly,
we observe that random motion of particles at early times
is not as important as random motion of particles at later times
for determining final locations at time $1$.
Motivated by this observation, we can define 
resampling times $0 < t_1 < \cdots < t_{K-1} < t_K = 1$
using the formula
\begin{align}
\label{resamplingtimes}
\int_0^{t_1} e^{2 \alpha t} \mathop{dt} = \int_{t_1}^{t_2} e^{2 \alpha t} \mathop{dt} = \cdots
= \int_{t_{K-1}}^{t_K} e^{2 \alpha t} \mathop{dt}
\end{align}
We can also define splitting functions $V_k\left(x\right) = C e^{\alpha \left(t_k-1\right)} \theta\left(x\right)$.
In this time-heterogeneous resampling strategy,
strength and frequency of splitting increase exponentially with time.
Splitting strength increases at the $1 \slash \alpha$ timescale with which
the OU process mean reverts to zero.
Splitting frequency increases at the $1 \slash \left(2 \alpha\right)$ timescale with which
the OU process variance reverts to $1$.
We note that a similar suggestion to increase the strength of splitting appears
in the work of Wouters and Bouchet \cite{wouters2016rare}.
The suggestion to increase the frequency of splitting is newly presented here.

Figure \ref{figure2} contrasts the different qualitative behavior of
time-homogeneous DMC, shown in red, and time-heterogeneous DMC, shown in blue.
In time-homogeneous DMC, the distribution of particles is immediately shifted toward high positions $x$ 
in the time interval $\left[0, 1 \slash \left(2 \alpha\right)\right]$;
in time-heterogeneous DMC, on the other hand,
the distribution of particles is shifted toward high positions $x$ at a later time interval
$\left[1 - 1 \slash \alpha, 1\right]$.

DMC can be used to estimate the probability 
$p = \Prob\left\{X_1 \geq U\right\}$
that the Ornstein-Uhlenbeck process exceeds a threshold $U$ at time $1$.
Following Definition \ref{def:DMC},
estimates take the form
\begin{equation}
\hat{p} = \frac{\overline{w}_{K-1}}{N} 
\sum_{i=1}^N \frac{\mathds{1}\left\{\xi_K^{\left(i\right)} \geq U\right\}}
{\exp\left\{V_{K-1}\left(\hat{\xi}_{K-1}^{\left(i\right)}\right)\right\}}
\end{equation}
We simulated rare extreme deviations of the Ornstein-Uhlenbeck process
ten thousand times
using DMC with a splitting intensity of $C = 2.5$.
We then computed estimates $\hat{p}$
and assessed error using the relative standard
deviation $\sqrt{\Var\left[\hat{p} / p\right]}$.
Results are shown in Table \ref{table1}.
The table shows for a range of $U$ values that time-heterogeneous DMC
is more accurate than time-homogeneous DMC.
Moreover, when studying the most extreme rare events,
time-heterogeneous DMC is more than fifty times more accurate than
direct sampling.
Thus, with direct sampling it would be necessary to increase the
sample size by a factor of more than a thousand
to obtain comparable error to time-heterogeneous DMC.

\begin{table}[!htbp]
\begin{tabular}{c|c|c|c}
~ & \shortstack[c]{direct \\ sampling \\ ~} & \shortstack[c]{time- \\ homogeneous \\ DMC} & \shortstack[c]{time- \\ heterogeneous \\ DMC} \\
\hline
$\Prob\left\{X_1 \geq 1\right\}$ & $0.073$ & $0.26$ & $0.096$ \\
$\Prob\left\{X_1 \geq 2\right\}$ & $0.21$ & $0.25$ & $0.11$ \\
$\Prob\left\{X_1 \geq 3\right\}$ & $0.90$ & $0.32$ & $0.17$ \\
$\Prob\left\{X_1 \geq 4\right\}$ & $6.1$ & $0.69$ & $0.37$ \\
$\Prob\left\{X_1 \geq 5\right\}$ & $67$ & $2.5$ & $1.3$
\end{tabular}
\caption{Relative standard errors.
With 1000 particles, time-heterogeneous DMC gives 
better estimates
for tail probabilities than time-homogeneous DMC
or direct sampling.}
\label{table1}
\end{table}

Having discussed the strengths of Diffusion Monte Carlo,
we now turn to a discussion of the method's shortcomings.
Under the best of conditions, Diffusion Monte Carlo is a highly effective rare event sampling strategy.
However, to sample a transformation of the OU process,
DMC can require delicate tuning,
which limits the method's practical appeal.

\begin{figure}[!htbp]
\includegraphics[width=8cm, trim = {0 0 0 .1cm}, clip]{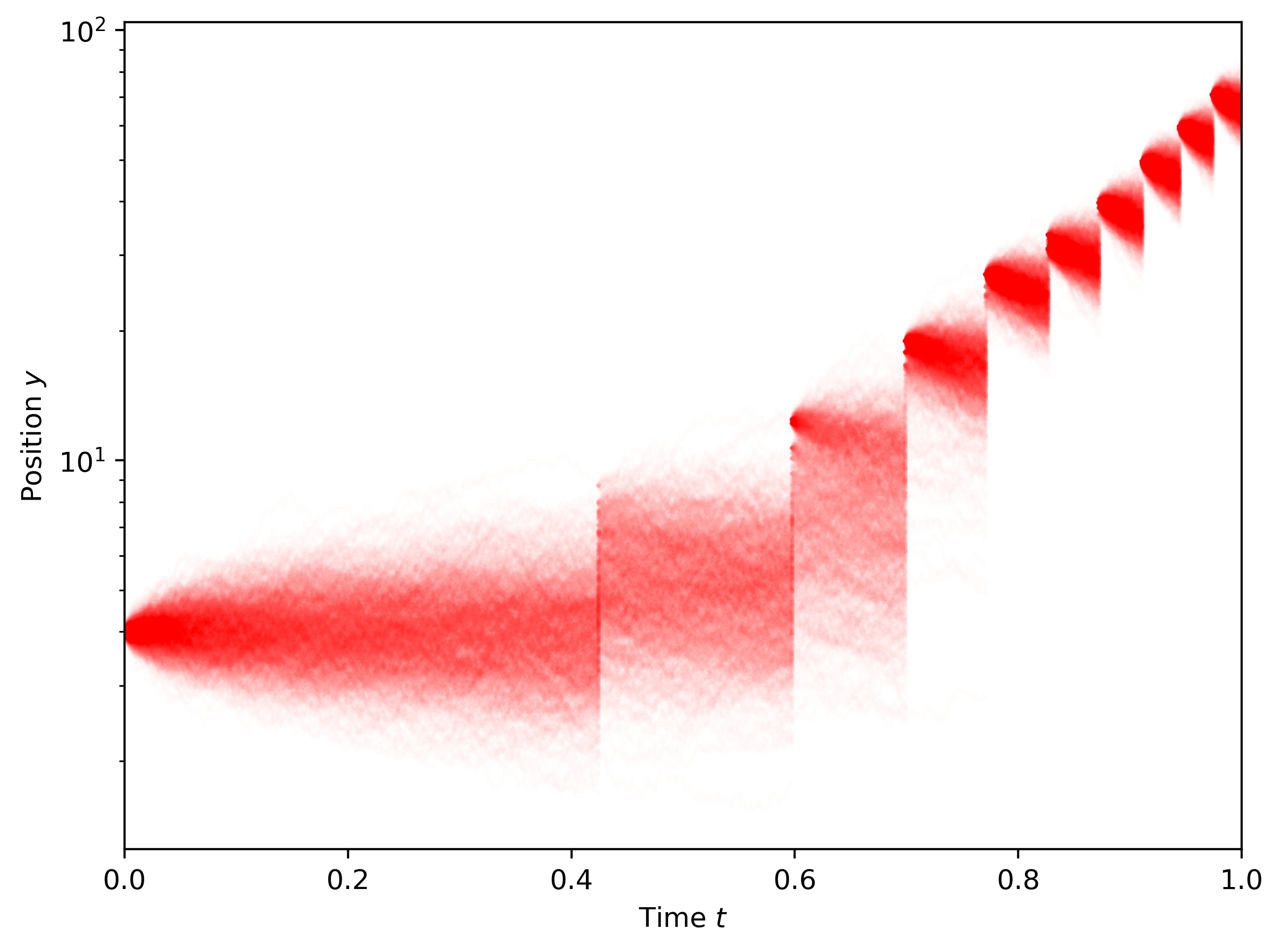}
\caption{With the wrong splitting functions $V_k$,
DMC can experience a catastrophic instability.
Extreme splitting leads to high correlations between particles
and error-prone estimates.}
\label{figure3}
\end{figure}

Consider using DMC to sample extreme deviations of the process
\begin{equation}
\begin{cases}
\mathop{d\log\left(Y_t \slash 4\right)} = -\alpha \log\left(Y_t \slash 4\right) \mathop{dt} + \sqrt{\alpha \slash 8} \mathop{dW_t} \\
Y_0 = 4
\end{cases}
\end{equation}
The process $Y_t$ is a nonlinear transformation of the OU process,
with $Y_t = 4 \exp\left\{X_t \slash 4\right\}$.
Figure \ref{figure3} illustrates what can happen when DMC is used
to sample $Y_t$ without a careful tuning of parameters.
We apply DMC with a reaction coordinate $\theta\left(y\right) = y$
and splitting functions
$V_k\left(y\right) = 2.5 e^{\alpha \left(t_k-1\right)} \theta\left(y\right)$.
The resulting DMC scheme performs well at the first resampling time, 
but as soon as the first particles reach positions $y > 10$, 
the algorithm becomes unbalanced.
Particles with the highest positions $y$ are split into dozens or hundreds of replicas.
Thus, particle positions $y$ become highly correlated,
leading to 
volatile and error-prone estimates for rare event probabilities.

While DMC can potentially be tuned to efficiently
sample the process $Y_t$,
the tuning process requires great care and flexibility.
In particular, no splitting function of the form 
$V_k\left(y\right) = C e^{\alpha \left(t_k - 1\right)} \theta\left(y\right)$
efficiently samples extreme deviations of $Y_t$.
A splitting function of a different parametric form is required.
Moreover, tuning becomes much more difficult when sampling a process with unknown dynamics.
Large amounts of data are necessary to tune DMC,
and gathering the necessary data from a complex model can be computationally expensive.


We draw two essential observations from the Ornstein-Uhlenbeck example.
First, DMC is most effective
when the strength and frequency of splitting increase over time.
Second, DMC is quite sensitive to the particular splitting functions $V_k$ that are used.
This sensitivity potentially compromises the real-world performance of DMC
and prompts the development of a more robust version of DMC.


\subsection{Quantile DMC}{\label{sub:QDMC}}

Quantile DMC is an elaboration of the DMC algorithm
algorithm, with additional adaptation steps
that make the scheme more robust.
The current section describes how adaptation steps are 
performed and explores the specific theoretical properties that 
explain the robustness of Quantile DMC.

The key difference between Quantile DMC and standard DMC
is that Quantile DMC adaptively rescales the reaction coordinate $\theta$
to match a target distribution $\nu_k$.
It is the rescaled reaction coordinate $\theta_k^{\prime}$
that is used for splitting and killing of simulations.

To perform Quantile DMC,
first define a series of resampling times $0 = t_0 < t_1 < t_2 < \cdots$.
For each resampling time $t_k$,
specify a target distribution $\nu_k$ for the rescaled reaction coordinate $\theta_k^{\prime}$
and specify $V_k^{\prime}\left(x\right)$,
a splitting function that increases with $\theta_k^{\prime}\left(x\right)$.
Quantile DMC begins with an initialization step and then
iterates over adaptation, reweighting, resampling, and mutation steps
according to the following definition:

\begin{definition}[Quantile DMC]{\label{def:QDMC}}
~
\begin{enumerate}
\item Initialization: Independently sample initial particles
$\xi_{0}^{\left(i\right)}\sim\Law\left(X_{0}\right)$ for $1 \leq i \leq N$
\item For $k=0,1,2,\ldots$,
\begin{enumerate}
\item Adaptation:
If $k = 0$, let $\gamma_0$ be a transport function from
$\frac{1}{N} \sum_{i = 1}^N \delta\left(\theta\left(\xi_0^{\left(i\right)}\right)\right)$
to $\nu_0$.
If $k > 0$, let $\gamma_k$ be a transport function from 
\begin{flalign}
&& \frac{\sum_{i = 1}^N \exp\left\{-V_{k-1}^{\prime}\left(\hat{\xi}_{k-1}^{\left(j\right)}\right)\right\} 
\delta\left(\theta\left(\xi_k^{\left(j\right)}\right)\right)}
{\sum_{i = 1}^N \exp\left\{-V_{k-1}^{\prime}\left(\hat{\xi}_{k-1}^{\left(j\right)}\right)\right\}}
\end{flalign}
to $\nu_k$.
Define the rescaled reaction coordinate $\theta_k^{\prime} = \gamma_k\left(\theta\right)$.
\item Reweighting: If $k = 0$, define initial weights
\begin{equation}
w_{0}^{\left(i\right)} = \exp\left\{V_0^{\prime}\left(\xi_0^{\left(i\right)}\right)\right\}
\end{equation}
If $k > 0$, define weights
\begin{flalign}
&& w_k^{\left(i\right)}
= \overline{w}_{k-1}
\exp\left\{V_{t_k}^{\prime}\left(\xi_k^{\left(i\right)}\right) - 
V_{t_{k-1}}^{\prime}\left(\hat{\xi}_{k-1}^{\left(i\right)}\right)\right\}
\end{flalign}
Define the average weight $\overline{w}_k = \frac{1}{N} \sum_{j=1}^N w_k^{\left(j\right)}$.
\item Resampling:
By splitting and killing particles $\left(\xi_k^{\left(i\right)}\right)_{1 \leq i \leq N}$,
create an ensemble of updated particles $\left(\hat{\xi}_k^{\left(j\right)}\right)_{1 \leq j \leq N}$
consisting of $N_k^{\left(i\right)}$ copies of each particle $\xi_k^{\left(i\right)}$.
The numbers $N_k^{\left(i\right)}$ are randomly chosen to satisfy
\begin{equation}
\label{condition}
\begin{cases}
\sum_{i=1}^N N_k^{\left(i\right)} = N \\
\E \left[N_k^{\left(i\right)}\right] = w_k^{\left(i\right)} \slash \overline{w}_k
\end{cases}
\end{equation}
\item Mutation: Independently sample
$\xi_{k+1}^{\left(i\right)}\sim \Law \left(X_{t_{k+1}}|X_{t_k}=\hat{\xi}_k^{\left(i\right)}\right)$
for $1\leq i\leq N$.
\end{enumerate}
\item Estimation: To approximate $\E \left[f\left(X_{t_k}\right)\right]$, 
Quantile DMC uses the estimate
\begin{equation}{\label{qdmcestimates}}
\E \left[f\left(X_{t_k}\right)\right]
= \frac{\overline{w}_{k-1}}{N} \sum_{i=1}^N \frac{f\left(\xi_k^{\left(i\right)}\right)}
{\exp\left\{V_{k-1}^{\prime}\left(\hat{\xi}_{k-1}^{\left(j\right)}\right)\right\}}
\end{equation}
\end{enumerate}
\end{definition}

Quantile DMC is distinguished from standard DMC by an adaptation step.
The adaptation step 
begins by estimating the distribution of
$\theta\left(X_{t_k}\right)$ using data from simulations.
At time $t_0$, the distribution of $\theta\left(X_0\right)$
is estimated using
$\eta_0 = \frac{1}{N} \sum_{i = 1}^N \delta\left(\theta\left(\xi_0^{\left(i\right)}\right)\right)$.
At later times,
the distribution of $\theta\left(X_{t_k}\right)$
is estimated using
\begin{equation}
\eta_k = \frac{\sum_{i = 1}^N \exp\left\{-V_{k-1}^{\prime}\left(\hat{\xi}_{k-1}^{\left(j\right)}\right)\right\} 
\delta\left(\theta\left(\xi_k^{\left(j\right)}\right)\right)}
{\sum_{i = 1}^N \exp\left\{-V_{k-1}^{\prime}\left(\hat{\xi}_{k-1}^{\left(j\right)}\right)\right\}}
\end{equation}

After estimating the distribution of $\theta\left(X_{t_k}\right)$, 
Quantile DMC builds a transformation
$\theta^{\prime}_k = \gamma_k\left(\theta\right)$
so that the distribution of
$\theta^{\prime}_k \left(X_{t_k}\right)$
approximates a target distribution $\nu_k$.
In particular, Quantile DMC builds a transport function \cite{rachev1998mass}
from $\eta_k$ to $\nu_k$ of the form
\begin{equation}
\gamma_k\left(x\right) = F_{\nu_k}^{-1}\left(F_{\eta_k}\left(x\right)\right)
\end{equation}
Here, $F_{\eta_k}$ is a distribution function for $\eta_k$, defined by
\begin{equation}
F_{\eta_k}\left(x\right) = \eta_k\left(\left(-\infty, x\right)\right) + \frac{1}{2}\eta_k\left(\left\{x\right\}\right)
\end{equation}
and $F_{\nu_k}^{-1}$ is a quantile function for $\nu_k$, defined by
\begin{equation}
F_{\nu_k}^{-1}\left(\alpha\right) = \inf\left\{x \in \mathbb{R}\colon F_{\nu_k}\left(x\right) \geq \alpha\right\}
\end{equation}
Since the transport function $\gamma_k$ maps the quantiles of $\eta_k$ to the quantiles
of $\nu_k$, we call this algorithm Quantile DMC.

\begin{figure}[!htbp]
\includegraphics[width=8cm, trim = {0 0 0 .1cm}, clip]{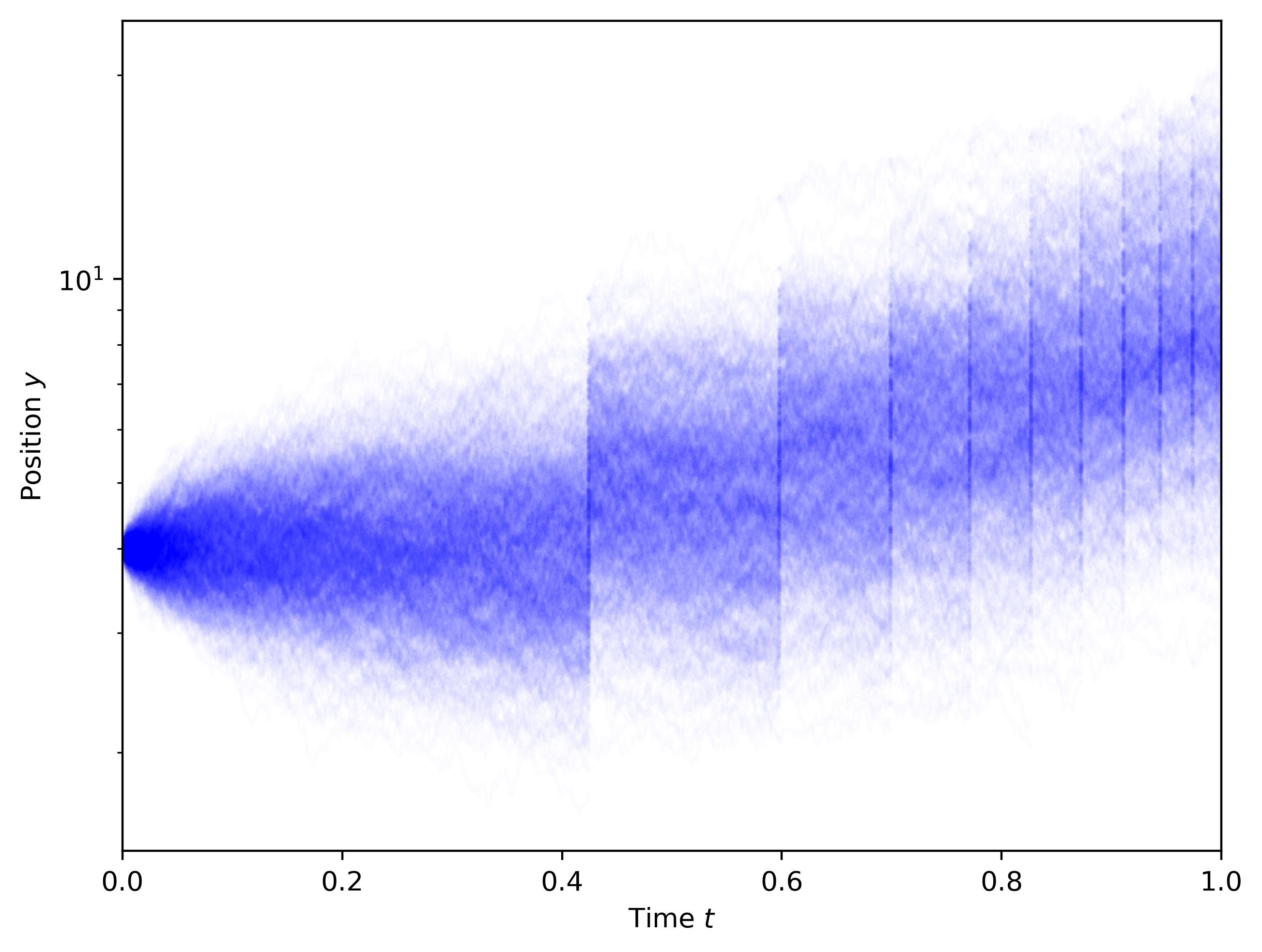}
\caption{Quantile DMC efficiently samples extreme deviations
of the nonlinear process
$\mathop{d\log\left(Y_t \slash 4\right)} = -2 \log\left(Y_t \slash 4\right) \mathop{dt} + 1 \slash 2 \mathop{dW_t}$.}
\label{figure4}
\end{figure}

An explicit example of Quantile DMC helps illustrate the main features of this new algorithm.
Consider using Quantile DMC to estimate the probability of extreme deviations of the process
\begin{equation}
\begin{cases}
\mathop{d\log\left(Y_t \slash 4\right)} = -\alpha \log\left(Y_t \slash 4\right) \mathop{dt} + \sqrt{\alpha \slash 8} \mathop{dW_t} \\
Y_0 = 4
\end{cases}
\end{equation}
This is the same nonlinear transformation of the OU process 
that was responsible for a catastrophic failure of DMC
in Section \ref{sub:example}.
To sample extreme deviations of the process $Y_t$, we define a reaction coordinate $\theta\left(y\right) = y$
and target distributions $\nu_k = N\left(0,1\right)$.
We use splitting functions
\begin{equation}
V_k^{\prime}\left(x\right) = 2.5 e^{\alpha\left(t_k - 1\right)} \theta^{\prime}_k\left(x\right)
\end{equation}
Figure \ref{figure4} presents results of these Quantile DMC simulations.
The behavior of Quantile DMC is highly stable.
Particles are nudged gently but forcibly in the direction of high $\theta$ values.
Explicit error calculations confirm that Quantile DMC is just as effective
at computing tail probabilities for this nonlinear transformation of the OU
process
as for the OU process itself.

The main advantage of Quantile DMC, compared to DMC,
is that Quantile DMC requires less tuning
when it is applied to a wide class of nonlinear processes.
This robustness is due to the fact that
Quantile DMC estimates have the same distribution
if the reaction coordinate $\theta$ is replaced with any other reaction coordinate $\tilde{\theta}$
that is a monotonic, one-to-one transformation of $\theta$.
Thus, for example, Quantile DMC is equally effective when sampling
from the OU process or from a nonlinear, montonic transformation of the OU process.
We note that this property of \emph{invariance under monotonic transformations} 
is also shared by Adaptive Multilevel Splitting \cite{cerou2007adaptive}
and Steered Transition Path Sampling \cite{guttenberg2012steered}
and thus appears to be an important property
underlying the success of a variety of effective rare event sampling methods.
We refer the reader to the appendix
for a proof of this theoretical property and further discussion
of convergence properties for Quantile DMC estimates.



\subsection{Implementation of Quantile DMC}{\label{sub:implementation}}

\begin{figure}[!htbp]
\includegraphics[width=8cm, trim = {0 0 0 .1cm}, clip]{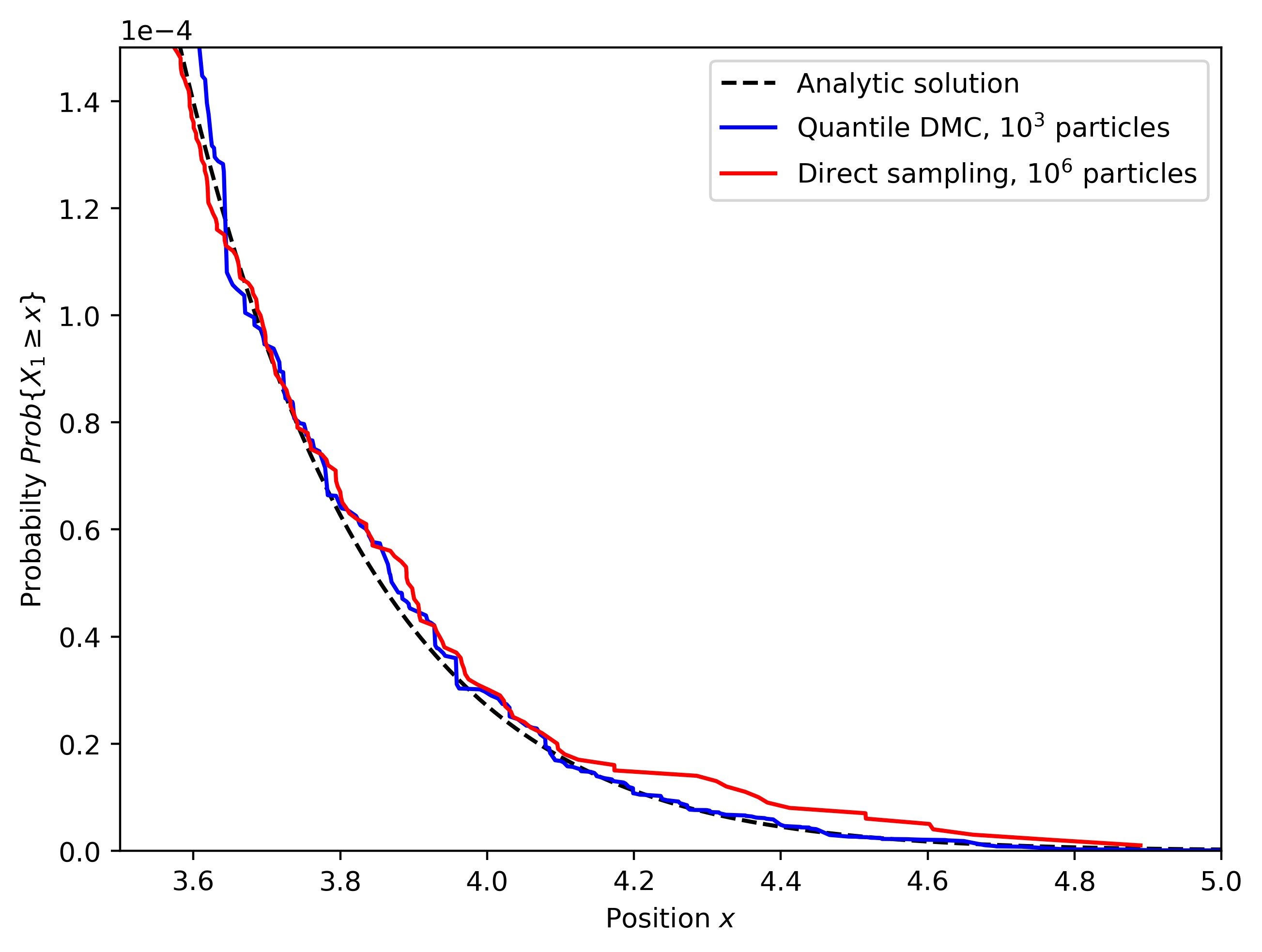}
\caption{Extreme tail probabilities obtained using Quantile DMC with $N = 1,000$ particles 
are more accurate than tail probabilities
obtained using direct sampling with $N = 1,000,000$ particles.}
\label{figure5}
\end{figure}

Quantile DMC
provides a method to estimate rare event probabilities with much reduced computational effort
compared to direct sampling.
For example,
Figure \ref{figure5} compares tail probabilities estimated using
Quantile DMC with $N = 1,000$ particles
to the same tail probabilities estimated using direct sampling
with $N = 1,000,000$ particles.
When sampling from the OU process,
Quantile DMC achieves better accuracy than direct sampling
but uses one thousand times less computational power.

Quantile DMC is straightforward to implement.
Whereas some rare event sampling algorithms
can require additional simulation time,
additional storage,
or additional manipulations of the underlying dynamical model
compared to direct sampling,
this is not the case with Quantile DMC.
The only additional cost of using Quantile DMC is the cost of resampling,
and the cost of resampling is often negligible compared to the cost of running complex simulations forward in time.
For the tropical cyclone simulations presented in Part \ref{sec:Extreme},
resampling was completed with ten lines of code
and a few seconds of processing time.

To maximize the efficiency of Quantile DMC, 
parameters should be adjusted
depending on the particular rare event sampling problem being investigated.
To illustrate this tuning process, 
we investigate optimal parameter choices for 
sampling from the OU process
with resampling times, target distributions, and splitting functions
\begin{equation}
\begin{cases}
\label{specifics}
\int_0^{t_1} e^{2 \alpha t} \mathop{dt} = \int_{t_1}^{t_2} e^{2 \alpha t} \mathop{dt} = \cdots
= \int_{t_{K-1}}^{t_K} e^{2 \alpha t} \mathop{dt} \\
\nu_k = N\left(0, 1\right) \\
V_k^{\prime}\left(x\right) = C e^{\alpha\left(1 - t_k\right)} \theta^{\prime}_k\left(x\right)
\end{cases}
\end{equation}
With this resampling schedule and this approach to splitting and killing particles, 
four parameters can impact the quality of Quantile DMC estimates:
the timescale parameter $\alpha$, the tilting constant $C$, the number of resampling times $K$,
and the number of particles $N$.

The simplest parameter to analyze is the timescale parameter $\alpha$.
Ideally, the parameter $\alpha$ should be the same timescale as 
the underlying Ornstein-Uhlenbeck process.
However, if the parameter $\alpha$ 
is over- or under-estimated by a factor of two,
we found in our experiments that
error increases by less than $20\%$.

To ensure Quantile DMC's effectiveness,
the tilting constant $C$ must be adjusted depending on the rareness
of the probabilities being investigated.
When estimating a tail probability
$\Prob\left\{X_1 \geq U\right\}$ for the OU process,
the optimal tilting constant $C$ lies within one or two units of $U$.
This numerical result is consistent with the fact that 
particles at time $t = 1$ are approximately normally distributed with mean $C$ and variance $1$,
and the $N\left(C, 1\right)$ distribution 
is a suitable importance sampling distribution 
for estimating $\Prob\left\{X_1 \geq U\right\}$
when $U$ lies close to $C$.

Parameters $K$ and $N$ should also be increased as probabilities being investigated
become rarer.
For example, when estimating $\Prob\left\{X_1 \geq 2\right\}$,
$100$ particles and $10$ resampling times
yield near-maximal efficiency.
When estimating $\Prob\left\{X_1 \geq 4\right\}$,
$1000$ particles and $100$ resampling times
are required for near-maximal efficiency.
In the second situation, the probability being estimated is rarer
and consequently more computational power is required.

\begin{figure}[!htbp]
\includegraphics[width=8cm, trim = {0 0 0 .1cm}, clip]{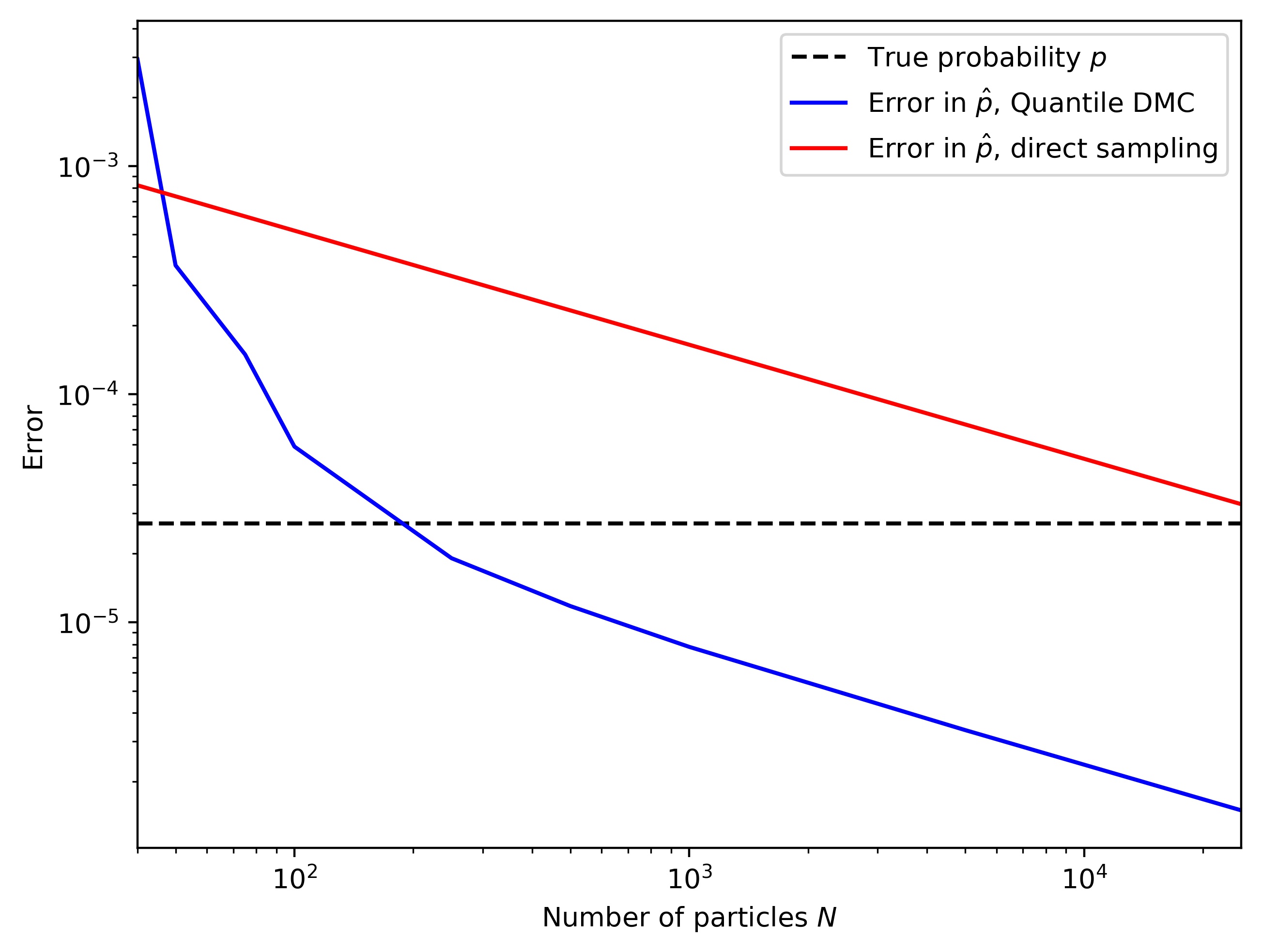}
\caption{Error in calculating $p = \Prob\left\{X_1 \geq 4\right\}$.
In Quantile DMC, error decays quickly for small $N$
and then levels off asymptotically to a $N^{-1 / 2}$ scaling.
In direct sampling, error decays with a perfect $N^{-1 / 2}$ scaling.}
\label{figure6}
\end{figure}

The efficiency gains from using Quantile DMC instead of direct sampling
become most dramatic when the number of particles $N$
exceeds a critical threshold.
Figure \ref{figure6} shows how Quantile DMC error 
decays quickly, at a faster than $N^{-1 / 2}$ rate as $N$ is increased from $10$ to $1000$.
Once the ensemble size reaches $N = 1000$ particles,
then error decreases less quickly, at an asymptotic $N^{-1 / 2}$ rate.


The tuning of parameters $\alpha$, $C$, $K$ and $N$
must be coupled with a careful selection of reaction coordinate $\theta$
to guarantee the effectiveness of Quantile DMC.
The goal of Quantile DMC is to emphasize paths leading to a rare event $A$
via splitting and to deemphasize paths leading away from $A$ through random killing.
Thus, the ideal reaction coordinate $\theta$ should anticipate what paths
lead to the rare event $A$.
One specific reaction coordinate that is appropriate for this goal 
is therefore the conditional probability function
\begin{equation}
\theta_k\left(x\right) = \Prob\left(\left.A\right| X_{t_k} = x\right)
\end{equation}
Here, the reaction coordinate $\theta_k$ changes
at each resampling time $t_k$.
This reaction coordinate is proven to be optimal
for a splitting algorithm similar to Quantile DMC \cite{cerou2006genetic}.

With Quantile DMC, identifying a reaction coordinate
can be easier than with DMC
because any monotonic, time-dependent transformation of $\Prob\left(\left.A\right| X_{t_k} = x\right)$
serves equally well as a reaction coordinate.
However, in a complex, high-dimensional system,
it can be challenging even to approximate
a monotonic transformation of $\Prob\left(\left.A\right| X_{t_k} = x\right)$.
In our analysis of tropical cyclone simulations in Part \ref{sec:Extreme},
we conclude that identifying an appropriate
reaction coordinate requires careful data analysis
and scientific insight into the model being simulated.

With an imperfect choice of reaction coordinate
Quantile DMC users should take care not to resample too often.
In our experiments with the OU process,
we can resample more than $K = 100$ times without any adverse effects
because the process of splitting/killing particles
is carefully tuned to the underlying dynamics.
However, for more complicated problems in which an ideal choice of $\theta$ is not available,
resampling should be performed as little as possible,
while sustaining the necessary particle dynamics\cite{liu2008monte}.
When testing a new reaction coordinate,
it is a good strategy to start with just a few
resampling times and increase resampling frequency
once the coordinate is proven to be effective.

In a complex, high-dimensional system, where the reaction coordinate $\theta$ is imperfect,
it is essential to gauge the quality of DMC estimates by providing error bars.
For estimates
\begin{equation}
\label{estimates}
\Prob\left\{X_{t_K} \in A\right\} \approx \frac{\overline{w}_{K-1}}{N}
\sum_{i=1}^N \frac{\mathds{1}\left\{\xi_K^{\left(i\right)} \in A\right\}}
{\exp\left\{V_{K-1}^{\prime}\left(\hat{\xi}_{K-1}^{\left(i\right)}\right)\right\}}
\end{equation}
it is possible to estimate variance $\sigma^2$
using\cite{chan2013general}
\begin{equation}
\label{advanced}
\frac{1}{N^2} \sum_{i=1}^N 
\left|\sum_{\anc\left(\xi_K^{\left(j\right)}\right) = i}
\frac{\overline{w}_{K-1} \mathds{1}\left\{\xi_K^{\left(j\right)} \in A\right\}}
{\exp\left\{V_{K-1}^{\prime}\left(\hat{\xi}_{K-1}^{\left(j\right)}\right)\right\}}
- \hat{p}\right|^2
\end{equation}
Here, $\anc\left(\xi_K^{\left(j\right)}\right)$ denotes
the ``ancestral index'' of particle $\xi_K^{\left(j\right)}$.
Tracing the ancestry of $\xi_K^{\left(j\right)}$
back to an initial particle $\xi_0^{\left(i\right)}$,
$\anc\left(\xi_K^{\left(j\right)}\right)$ is the index $i$
for the initial particle.
We note that the variance estimator $\hat{\sigma}^2$ 
is most accurate when the sample size $N$ is very high
and when the number of resampling times $K$ is small,
which is not always the case in practical simulations.
Properties of the variance estimator $\hat{\sigma}^2$ are discussed at more length in the appendix.

In summary, Quantile DMC is straightforward to implement
and becomes increasingly effective relative to other sampling methods
as parameters of splitting/killing are tuned and as the number of particles $N$ increases.
However, Quantile DMC can perform poorly when the reaction coordinate $\theta$
fails to anticipate paths leading to rare event states.
Lastly, we have shown how to provide rough 
error bars that assess the accuracy of Quantile DMC estimates.


\section{Extreme mesoscale weather}{\label{sec:Extreme}}

In this section we discuss why it is difficult to estimate the frequency of extreme 
mesoscale weather and how rare event sampling methods like Quantile DMC
can potentially assist in calculations.
Then we focus attention on the frequency of intense tropical cyclones
and present simulations which illustrate both the potential benefits
of rare event sampling and the work that remains to be done.


\subsection{Frequency of extreme weather}

While extreme weather events
such as heat waves, floods, and tropical cyclones are rare,
they can cause immense damage and fatalities
\cite{pielke2000precipitation, pielke2008normalized, coumou2012decade}.
Understanding the frequency of extreme weather
is therefore an essential task,
both for real-world disaster preparedness
and for assessing weather's impact on society.
The study of weather extremes is more relevant than ever,
since evidence points to changing frequencies
of extreme weather events with climate change 
\cite{knutson2010tropical, pall2011anthropogenic, coumou2012decade}.

To understand the frequency of extreme weather,
observations provide the most fundamental data source,
but this data has important limitations.
For many potential extreme weather events
no historical analogue exists.
Storms can occur in surprising places.
Droughts can afflict new areas.
Even when historical data are available,
measurements can be sparse and sometimes corrupted \cite{saha2010ncep}.
Most critically, as the climate changes,
the frequency and intensity of tropical cyclones, of heat waves, and of flooding
are expected to change,
so that historical measurements will become less relevant
\cite{knutson2010tropical, pall2011anthropogenic, coumou2012decade}.
For all these reasons, climate simulations 
provide essential additional insight into
extreme weather in historical, current, and potential future climates.

When modeling extreme weather,
there is a tradeoff between bias and variance.
While inexpensive models can be run many times, leading to 
low-variance estimates, 
these estimates can be highly biased.
More computationally intensive models potentially provide a less biased
climatology of extreme weather 
\cite{hirabayashi2008global, pall2011anthropogenic, davis2008prediction}
yet these models cannot be run for as long
or with as many ensemble members,
due to limited computational resources.
Because of these practical limitations,
estimates can have high variance,
particularly for the rarest, most extreme weather events.


Mesoscale extreme weather, 
such as floods and tropical cyclones,
provides a stark example of the need for increased statistical accuracy without
sacrificing model fidelity.
Because mesoscale extreme weather occurs on a smaller spatial scale (10-1000km)
at which simplifying assumptions for thermodynamics and dynamics begin to fail,
running accurate mesoscale weather models can be enormously expensive.
It is here that the computational burden is the greatest.
It is here, therefore, that rare event sampling methods stand
to provide the greatest benefit.

In extreme mesoscale weather simulations
where rare event sampling could potentially provide a benefit,
a process known as ``dynamical downscaling" \cite{maraun2010precipitation} is now common.
First a Global Climate Model (GCM) simulates a coarse-resolution version of an extreme weather event.
Using initial and boundary conditions from the GCM,
a high-resolution regional model with more complex physics
then enhances the GCM output,
simulating the local details of the extreme weather event.

The dynamical downscaling approach is necessary because current GCMs 
cannot simulate the details of mesoscale weather
that are essential for damage assessment.
For example, the peak winds of a tropical cyclone
are underestimated by a GCM \cite{bender2010modeled}
and a GCM cannot resolve the overflowing riverbeds 
that lead to flooding \cite{hirabayashi2008global, pall2011anthropogenic}.
Versions of dynamical downscaling have become standard in storm surge modeling \cite{lin2016grey}, 
flood modeling \cite{pall2011anthropogenic}, 
and tropical cyclone modeling \cite{bender2010modeled}.

A simple probabilistic interpretation helps clarify
how dynamical downscaling can estimate
the probability of an extreme weather event occurring.
Let $A$ denote an extreme event
and let $B$ denote the coarse-scale meteorological conditions
that are necessary for event $A$ to occur. 
We can then write,
\begin{multline}
\Prob\left\{\text{event } A\right\} = \Prob\left\{\text{conditions } B\right\} \\
\times \Prob\left\{\left.\text{event } A \, \right| \text{conditions } B\right\}
\end{multline}
A GCM is used to evaluate the first probability $\Prob\left\{\text{conditions } B\right\}$;
whereas, a regional model is used to evaluate the second probability
$\Prob\left\{\left.\text{event } A \, \right| \text{conditions } B\right\}$
using output from the GCM.
For example, Bender and coauthors \cite{bender2010modeled}
identified proto-cyclones in a GCM
and then used a high-resolution regional model to simulate
the intensity evolution of proto-cyclones into full-fledged tropical cyclones.

Applying rare event sampling in a dynamical downscaling context
is a multi-tiered process.
A GCM can be run either directly
or with a splitting method such as Quantile DMC.
Then, starting from the intial conditions selected from GCM output,
a regional weather model can be run either directly
or with a splitting method such as Quantile DMC.
At multiple stages of the dynamical downscaling process,
statistics can be potentially improved by a judicious application of rare event sampling.

In summary, there is a pressing need for rare event sampling of mesoscale extreme weather,
since mesoscale simulations are enormously expensive.
The incorporation of rare event sampling 
into extreme weather calculations can
enable higher-resolution, more computationally intensive models,
ultimately leading to more accurate extreme weather risk assessment.


\subsection{Tropical cyclone test case}


\subsubsection{Motivation for simulations}

Tropical cyclones rank among the deadliest natural disasters in human history.
Approximately 300,000 died in the 1970 Bhola cyclone,
and an estimated 138,000 died in the 2008 cyclone Nargis \cite{emanuel1988toward, fritz2009cyclone}.
While high-intensity tropical cyclones (TCs) are rare,
they are the most destructive and fatal TCs \cite{bakkensen2016risk, pielke2008normalized}.
Moreover, the frequency of the most intense storms
is expected to increase with climate change,
the precise rate of change being an open area of research \cite{knutson2010tropical}.
Understanding the upper tail of intensity for TCs
is therefore of paramount societal concern.

Reducing computational cost is
a central priority for the TC modeling community.
TCs are most accurately simulated using high-resolution weather models with
1-10km horizontal resolution \cite{davis2008prediction}.
High spatial resolution is required to resolve the storm eyewall,
where winds are the strongest.
Increasing horizontal resolution can lead to more accurate simulations;
however, increased resolution comes at a steep computational cost.
Doubling horizontal resolution requires an eight-fold increase in computational expense
because resolution must be doubled in the zonal and meridional directions, 
and the timestep must be cut in half to ensure numerical stability.

Helping to alleviate the computational burden of TC modeling,
rare event sampling potentially provides a means
to accurately estimate TC statistics with a reduced sample size
of high-resolution simulations.
Here, as a proof of concept, we apply Quantile DMC to 
estimate statistics for two high-intensity TCs.
For initial and boundary conditions,
we use reanalysis data for two storms that achieved Category 4 status in the real world
but did not achieve the highest intensity level, Category 5 status.
We model these storms using a stochastic model
that predicts a range of possible intensities,
different from the real-world intensities of Hurricane Earl
and Hurricane Joaquin.
Starting from coarse-scale proto-cyclones, 
we nudge the evolution of storms toward high intensities using the Quantile DMC algorithm
and then estimate the probability of high-intensity manifestations 
of these storms.
We then compare the efficiency of Quantile DMC
to the efficiency of direct sampling for this estimation problem.

These simulations are envisioned as a first step toward
the goal of using Quantile DMC to study 
the probability of intense TCs in historical,
present, and future climates.
In our simulations,
we use Quantile DMC to provide statistics for just two storms.
In the future, however,
as Quantile DMC is applied to study the frequency of intense TCs
in different climates,
it will be necessary to start simulations from an ensemble of hundreds
of different proto-cyclones.
For these initial investigations, we use a small sample size $N = 100$,
whereas a sample size of $N = 1000$ would be more appropriate
for a full implementation of Quantile DMC in the future.


\subsubsection{Simulation details}

The 2010 storm Earl was a long-lived hurricane of tropical origin that
came very close to the Eastern seaboard of the United States but ultimately did not make landfall,
passing 150km off the coast of Massachusetts.
The 2015 storm Joaquin was a hurricane
of extratropical origin that intensified more rapidly than expected,
leading to the worst U.S. maritime disaster in decades,
the sinking of the cargo ship El Faro
with all 33 sailors aboard \cite{langewiesche2018clock}.

We model Hurricanes Earl and Joaquin
using the Advanced Research Weather Research and Forecasting Model \cite{skamarock2008description}
(ARW, version 3.9.1.1), 
which has been applied extensively for hurricane research in the past \cite{davis2008prediction}.
ARW is a finite-difference model 
whose governing equations include
energy conservation,
mass conservation,
a thermodynamic law, and an equation of state.
ARW incorporates parametrized physics schemes for
precipitation processes, heat/moisture fluxes over land,
radiation, and mixing in atmospheric columns.
ARW uses artificial dissipation and filters
to achieve numerical stability.

We have previously used the ARW model to simulate
intense tropical cyclones \cite{plotkin2018maximizing}; 
however, our previous simulations did not incorporate the Quantile DMC algorithm.
In \citet{plotkin2018maximizing},
we used an optimization strategy
to identify a maximum likelihood pathway
for a simulated tropical cyclone to achieve a high intensity at a terminal time.
While the optimization approach is useful for identifying
factors that can cause tropical cyclones to intensify,
it is less suitable for computing tropical cyclone statistics.
Statistics of TCs can depend on myriad possible paths, 
and sampling these paths is a necessary requirement
for accurate estimation of statistics.
Thus, in the current work, we present Quantile DMC as an additional tool, 
uniquely suited to low-variance calculation of tropical cyclone statistics.

Among weather models, ARW has the advantage that it supports
vortex-following nested domains.
Three domains of different resolutions are often
used in TC simulations. 
The outer domain is static,
while the inner domains follow a local minimum in the 500hPa geopotential height field,
indicating a TC's location.
Vortex-following domains enable high resolution around the eye of the storm
without the computational expense of high resolution across the entire storm path.

Our simulations use a timestep of 6.7s and horizontal resolution of 2km in the inner domain,
a timestep of 20s and resolution of 6km in the middle domain, 
and a timestep of 60s and resolution of 18km in the outer domain.
The inner domain stretches 468km x 468km,
the middle domain 1404km x 1404km,
and the outer domain 5382km x 5382km.
All domains use 40 vertical levels.
The physics parametrizations are the same as used by 
Judt, Chen, and Berner \cite{judt2016predictability}.
Convection is explicitly simulated in the two inner domains
whereas convection is parametrized in the outermost domain.

Simulations are randomly perturbed using the
Stochastic Kinetic Energy Backscatter (SKEB) scheme \cite{berner2011model}.
Perturbations from this physics scheme are smooth in space and time,
but they change rapidly, modeling the effects of small-scale turbulent processes.
Potential temperature and the non-divergent component of horizontal wind
are both independently perturbed with forcing terms
\begin{equation}
F\left(x, y, z, t\right) = \sum_{j,k} \Re\left\{F_{j,k} \left(t\right) H_{j,k}\left(x,y\right) e^{i C_{j,k,z}} \right\}
\end{equation}
where $H_{j,k}$ are the Fourier modes for the domain and
$C_{j,k,z}$ are constants that produce a westward phase tilt in the perturbation field.
$F_{j,k}$ terms evolve randomly, according to a complex-valued Ornstein-Uhlenbeck process
with decorrelation timescale $\alpha^{-1} \approx 0.5\text{hrs}$, that is,
\begin{equation}
\mathop{dF_{j,k}} = -\alpha F_{j,k} \mathop{dt} + \sqrt{2 \alpha \sigma^2_{j,k}} \mathop{dW_{j,k}}
\end{equation}
where $W_{j,k}$ denote independent complex-valued Brownian motions and
the noise amplitude $\sigma^2_{j,k}$ decreases as a power law of $\sqrt{j^2 + k^2}$.
The SKEB scheme is implemented as a physics module within the ARW software.
The paper of \citet{berner2011model} describes how this physics module
discretizes the underlying OU process dynamics as a first-order autoregressive process.
At every ARW timestep, the perturbation field is updated, and perturbations
provide a series of small, frequent changes influencing model dynamics.

In our previous simulations of tropical cyclones
\cite{plotkin2018maximizing},
we used an alternative probabilistic model for ARW perturbations,
different from the SKEB scheme.
In \citet{plotkin2018maximizing}, 
a Gaussian perturbation is applied once per hour to the ARW model.
The optimization strategy used in \citet{plotkin2018maximizing}
would not have been practical to apply to the SKEB scheme.
The SKEB scheme perturbs simulations many times per minute,
making a derivative-based optimization challenging,
but leading to small and physically realistic perturbations.

We use direct sampling and Quantile DMC to sample extreme intensities of
Earl and Joaquin
with an ensemble size of $N = 100$ simulations.
Direct sampling runs are seven-day forward runs of the ARW model with SKEB.
Quantile DMC runs are seven-day forward runs, which incorporate splitting and killing of trajectories.
Since the ARW model with SKEB is a Markovian model,
Quantile DMC provides unbiased estimates
of a wide range of statistics of tropical cyclones,
including pathways, characteristics, and frequencies.
For simplicity, however,
this paper discusses only the probability of intense tropical cyclones occurring. 

\begin{table}[!htbp]
\begin{tabular}{l|l|l|l}
~ & ~ & Earl & Joaquin \\
\hline
$t_0$ & start & August 27 00:00 & September 29 00:00 \\
$t_1$ & resample 1 & August 30 00:00 & October 1 00:00 \\
$t_2$ & resample 2 & August 31 00:00 & October 2 00:00 \\
$t_3$ & resample 3 & August 31 12:00 & October 2 12:00 \\
$t_4$ & resample 4 & September 1 00:00 & October 3 00:00 \\
$t_5$ & end & September 3 00:00 & October 6 00:00
\end{tabular}
\caption{Start times, end times, and resampling times for Earl and Joaquin (all time zones are UTC).}
\label{table2}
\end{table}

Table \ref{table2} describes the specific start and end times for the simulations.
Start times are selected near the beginning of the hurricane life cycle.
End times are selected seven days after start times,
which gives hurricanes sufficient time to reach peak intensity
and then recede for at least two days.
For the Quantile DMC runs,
we select a time $T$ when we expect each storm to achieve peak intensity.
We resample at times $t_1 = T - 48hr$, $t_2 = T - 24hr$, $t_3 = T - 12hr$ and $t_4 = T$.
Thus, resampling increases in frequency
in an attempt to maximize Quantile DMC efficiency.
Resampling stops once hurricanes are expected to achieve peak intensity,
since resampling too often can increase the variance of Quantile DMC estimates.
The specific time $T$ for each storm is identified using official best track historical data \cite{landsea2013atlantic}.

The parameters for our Quantile DMC simulations
are defined as follows:
\begin{enumerate}
\item
The reaction coordinate $\theta$ is the deviation of sea surface pressure from a hydrostatically-balanced reference
state \cite{skamarock2008description} at the storm core.
\item
We assume the reaction coordinate $\theta$ can be modeled as a monotonic,
time-dependent transformation of an OU process.
Since  $N\left(0, 1\right)$ distributions work well as target distributions
when sampling from the OU process,
our target distribution is $N\left(0, 1\right)$ at each adaptation step.
\item
Splitting functions take the form $V_k^{\prime} = C e^{\alpha\left(t - T\right)} \theta^{\prime}_k$.
The decorrelation timescale $\alpha^{-1} = 3d$ is the appropriate timescale for large-scale differences
to emerge in TC development in the ARW model \cite{judt2016predictability}.
The splitting constant $C = 1$ is
appropriate for estimating intensity quantiles up to the $99.9$th percentile of intensity.
Moreover, when sampling an OU process with a splitting constant of $C = 1$, 
four rounds of resampling and a sample size of $N = 100$ simulations are sufficient to ensure 
the effectiveness of Quantile DMC
compared to direct sampling.
\end{enumerate}

For both Quantile DMC and direct sampling, 
equivalent computing resources are required
on the University of Chicago Research Computing Center high-performance cluster:
100 nodes ran continuously for 2 days with 28 CPUs per node and 2 gigabytes of RAM per CPU.


\subsubsection{Simulation results}{\label{subsub:Results}}

\begin{figure*}
\includegraphics[width=15cm, trim = {0 0 0 .1cm}, clip]{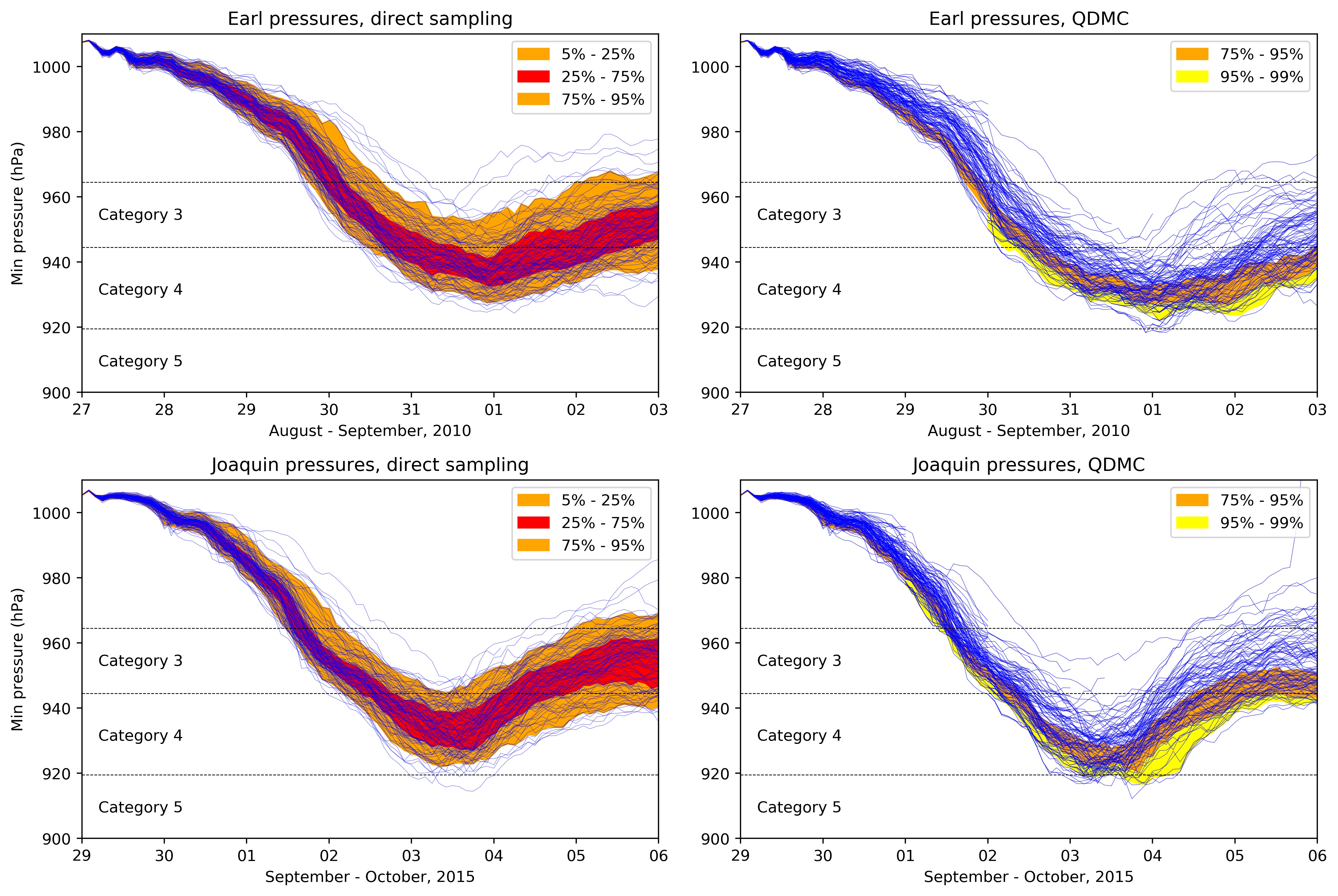}
\caption{Intensity trajectories for $N = 100$ direct sampling and Quantile DMC simulations
with intensity quantiles estimated from data.
In direct sampling (left), trajectories occupy the middle
quantiles of intensity (orange and red);
in Quantile DMC (right), trajectories are more likely to occupy
extreme quantiles of intensity (yellow).
While direct sampling trajectories span from the beginning to the end of simulations,
Quantile DMC trajectories can end when killing occurs.}
\label{figure9}
\end{figure*}

In Figure \ref{figure9}, we present intensity trajectories
for direct sampling and Quantile DMC runs
for Hurricane Earl and Hurricane Joaquin.
Direct sampling
trajectories typically occupy the middle quantiles of intensity,
whereas Quantile DMC
trajectories are more likely to occupy the upper quantiles of intensity.
For example, direct sampling produces zero Category 5 realizations of Earl
and only four of Joaquin.
In contrast, Quantile DMC produces three Category 5 realizations of Earl 
and 22 of Joaquin.
Therefore Quantile DMC is successful at simulating more intense storms compared
to direct sampling.

When presenting results,
we measure TC intensity using minimum sea surface pressure.
While it is also common to see TC intensity reported 
using maximum wind speed,
econometric analysis finds that minimum sea surface pressure is a better predictor of TC damage and fatalities
than is wind speed \cite{bakkensen2016risk}.
Pressure combines information on wind speed 
and storm size \cite{chavas2017physical},
thereby giving a more holistic indication of TC damage \cite{zhai2014dependence}.
We note that historically the Saffir-Simpson hurricane scale
combined maximum wind speed, minimum pressure and maximum storm surge information to 
classify TCs into Categories 1 (least intense) to 5 (most intense).
More recently, the scale was renamed the Saffir-Simpson hurricane wind scale,
and Categories 1 - 5 are defined by maximum wind speeds alone \cite{ofcm2010national}.
When presenting results, we use the historical Saffir-Simpson hurricane scale to define Categories 1 - 5 in terms of
minimum sea surface pressure.

%

We can use data from Quantile DMC and
direct sampling runs to estimate a range of statistics
associated with storm intensity.
In particular, we estimate cumulative distribution functions
for random variables $P_T$ and $P_{life}$.
$P_T$ is TC intensity at the particular time $T$
when each storm is expected to reach maximum strength,
namely, September 1 00:00 UTC for Earl and October 3 00:00 UTC for Joaquin. 
$P_{life}$ is the strongest TC intensity over the entire seven days of simulated time. 
The cumulative distribution function for $P_T$ is defined by
$F\left(U\right) = \Prob\left\{P_T \leq U\right\}$, where $U$ ranges over all possible pressures.
To estimate $\Prob\left\{P_T \leq U\right\}$
from Quantile DMC data,
we use
\begin{equation}
\Prob\left\{P_T \leq U\right\}
\approx \frac{\overline{w}_{K-1}}{N} \sum_{i=1}^N 
\frac{\mathds{1}\left\{P_T\left(\xi_K^{\left(i\right)}\right) \leq U\right\}}
{\exp\left\{V_{K-1}^{\prime}\left(\hat{\xi}_{K-1}^{\left(j\right)}\right)\right\}}
\end{equation}
To estimate $\Prob\left\{P_T \leq U\right\}$ from direct sampling data,
we use
\begin{equation}
\Prob\left\{P_T \leq U\right\}
\approx \frac{1}{N} \sum_{i=1}^N \mathds{1}\left\{P_T\left(\xi^{\left(i\right)}\right) \leq U\right\}
\end{equation}
We apply analogous formulas when estimating the cumulative distribution function for $P_{life}$.

\begin{figure*}
\includegraphics[width=15cm, trim = {0 0 0 .1cm}, clip]{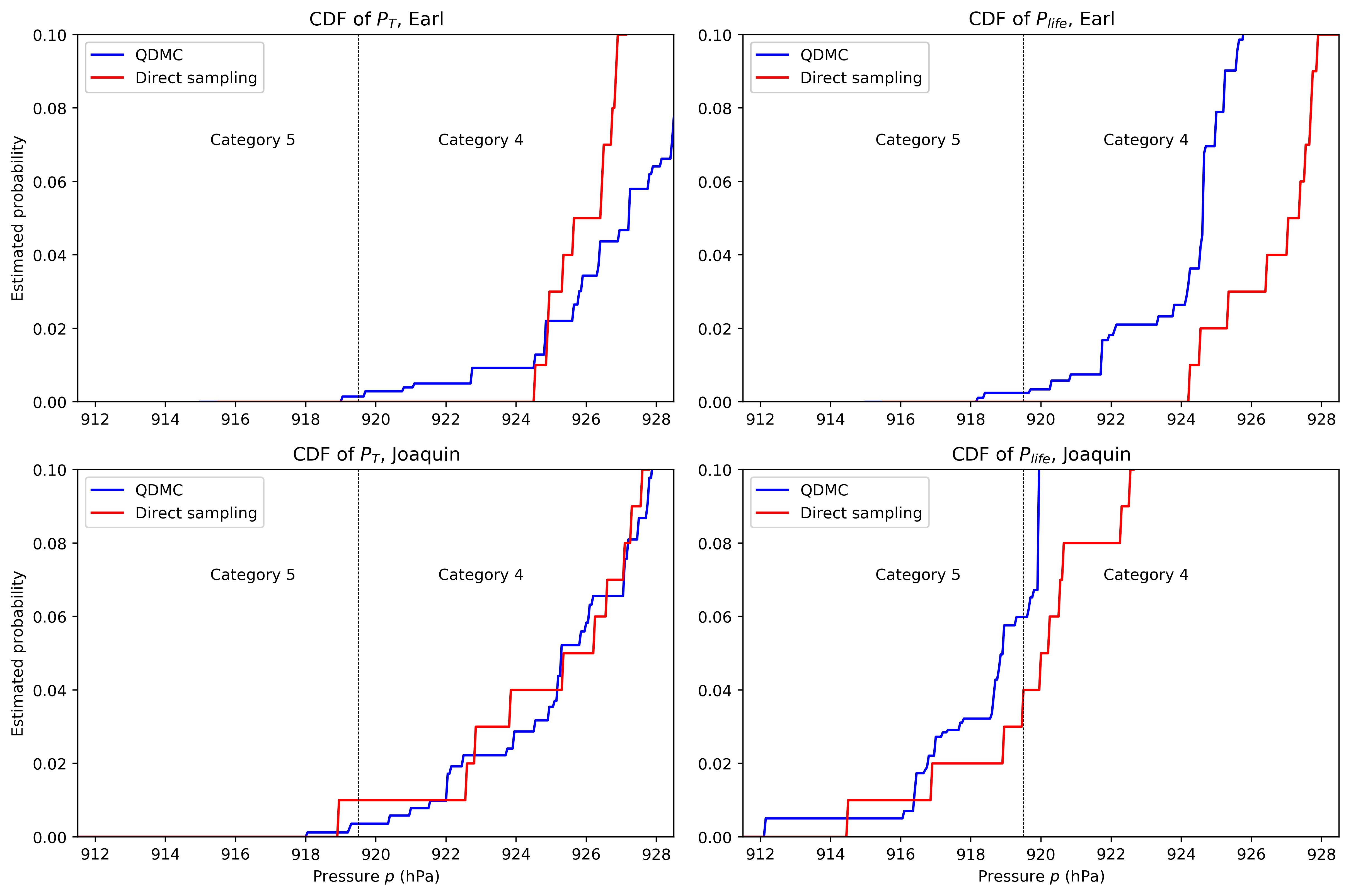}
\caption{Estimates of cumulative distribution functions $\Prob\left\{P_T \leq p\right\}$ and 
$\Prob\left\{P_{life} \leq p\right\}$
are roughly consistent between Quantile DMC and direct sampling.
However, estimates from direct sampling exhibit
large jumps, which may be a sign of higher error.}
\label{figure10}
\end{figure*}

Figure \ref{figure10} provides side-by-side comparisons
between the direct sampling and Quantile DMC estimates.
Intensity estimates from Quantile DMC are statistically consistent with intensity estimates
from direct sampling.
For example, the estimated probability for Earl to reach Category 5 status
is 0\% from direct sampling and 0.2\% from Quantile DMC.
The estimated probability for Joaquin to reach Category 5 status
is 4\% from direct sampling and 6\% from Quantile DMC.
The disagreement between estimates falls well within the range of random error,
particularly if one or the other estimate has high relative variance.
The cumulative distribution functions estimated from direct sampling 
exhibit sharper jump discontinuities
compared to the relatively smooth behavior of the CDFs from Quantile DMC.
This jumpy behavior may reflect a greater degree of error in the direct sampling estimates
if it can be assumed that the distribution of hurricane pressures is smooth.

To check the hypothesis that Quantile DMC estimates are more accurate,
we can use data from Quantile DMC and direct sampling runs to gauge
the variance in our estimates.
For Quantile DMC estimates of $\Prob\left\{P_T \leq U\right\}$,
we assess variance using
\begin{multline}
\hat{\sigma}^2 = \\
\frac{1}{N^2} \sum_{i=1}^N 
\left|\sum_{\anc\left(\xi_K^{\left(j\right)}\right) = i}
\frac{\overline{w}_{K-1} \mathds{1}\left\{P_T\left(\xi_K^{\left(i\right)}\right) \leq U\right\}}
{\exp\left\{V_{K-1}^{\prime}\left(\hat{\xi}_{K-1}^{\left(j\right)}\right)\right\}}
- \hat{p}\right|^2
\end{multline}
For direct sampling estimates of $\Prob\left\{P_T \leq U\right\}$,
we assess variance using $\hat{\sigma}^2 = \frac{1}{N}\hat{p}\left(1 - \hat{p}\right)$.

\begin{figure*}
\includegraphics[width=15cm, trim = {0 0 0 .1cm}, clip]{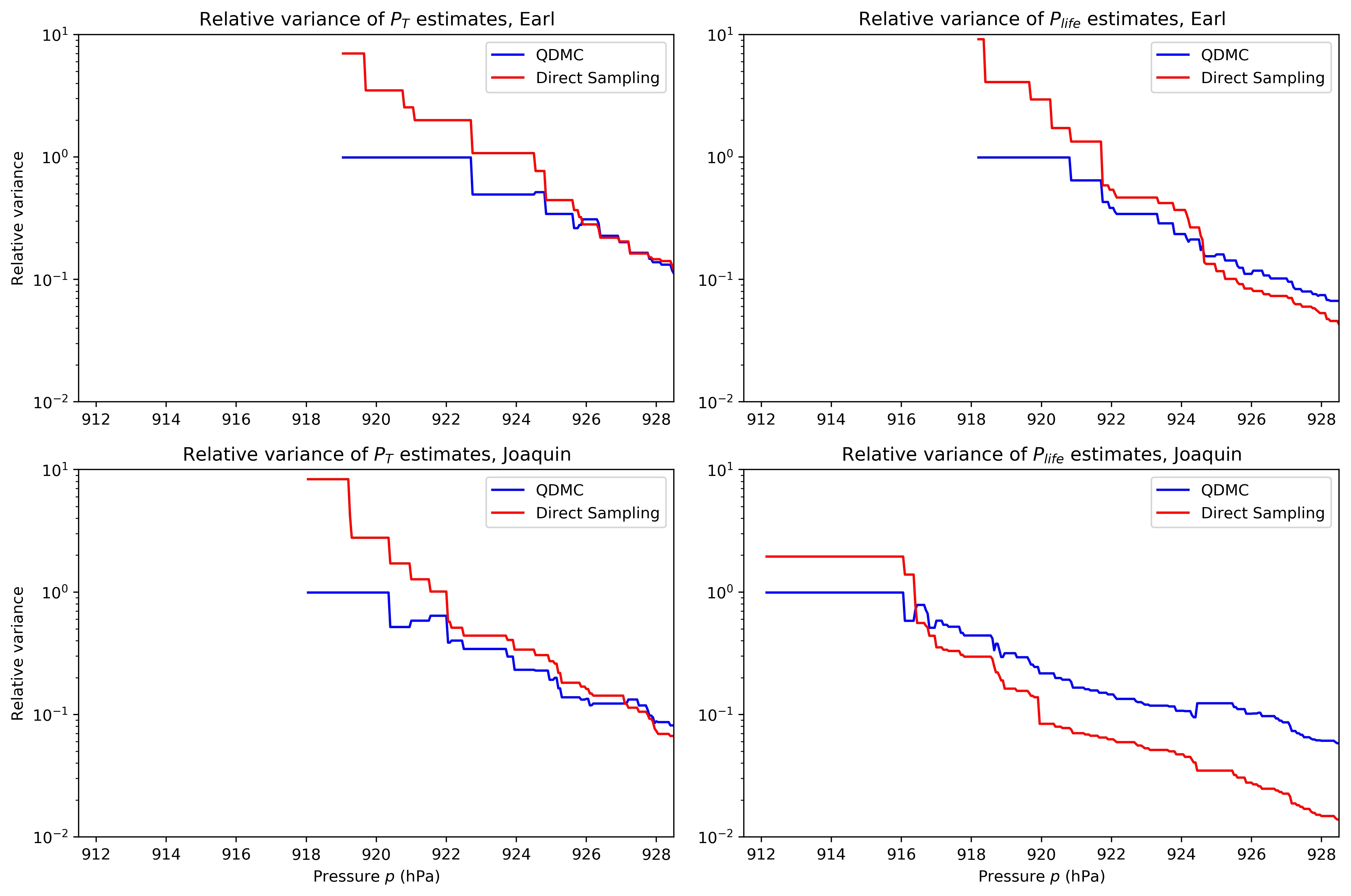}
\caption{Variances are lower for Quantile DMC than direct sampling
at extreme tail pressures $p$.
Quantile DMC provides the greatest benefit when estimating $P_T$ statistics
for Earl and Jaquin
and $P_{life}$ statistics for Earl.
Quantile DMC provides less of an improvement when estimating $P_{life}$ statistics for Joaquin.}
\label{figure11}
\end{figure*}

Figure \ref{figure11} provides side-by-side comparisons 
between direct sampling and Quantile DMC relative variances $\hat{\sigma}^2 \slash \hat{p}^2$.
For many important rare event sampling estimates,
Quantile DMC provides substantial variance reduction compared to direct sampling.
When estimating the $P_T$ distribution, 
Quantile DMC gives reduced variance for 
all pressures lower than 925hPa for both Earl and Joaquin.
When estimating the $P_{life}$ distribution,
Quantile DMC gives reduced variance for all
pressures lower than 925hPa for Earl and 916hPa for Joaquin.
At the lowest pressures,
the variance of Quantile DMC is two
to ten times lower than the variance of direct sampling.

An important reason for presenting the $P_T$ distribution is to show that 
Quantile DMC is highly effective at sampling intense storms at the reference time $T$. 
More generally, Quantile DMC is effective at 
sampling intense storms in the twelve hours leading up to the reference time $T$ 
and the twelve hours following the reference time $T$.
Outside this window of time, Quantile DMC may be less effective at sampling intense storms.
It is therefore of central importance to make sure the reference time $T$
aligns with the time of maximum intensity.
To achieve this goal, it may be necessary to run the model in advance of Quantile DMC simulations
or at least predict based on initial and boundary conditions when the model will achieve peak intensity.

The one anomalous result in the pattern of variance reduction due to Quantile DMC
is Quantile DMC's limited benefit
when estimating the $P_{life}$ distribution for Hurricane Joaquin.
To shine light on the limitations of Quantile DMC for this particular
estimation problem,
we can examine the family weights 
\begin{equation}
W^{\left(i\right)} = \sum_{\anc\left(\xi_K^{\left(j\right)}\right) = i}
\frac{\overline{w}_{K-1} \mathds{1}\left\{P_{life}\left(\xi_K^{\left(i\right)}\right) \leq U\right\}}
{\exp\left\{V_{K-1}^{\prime}\left(\hat{\xi}_{K-1}^{\left(j\right)}\right)\right\}}
\end{equation}
contributing to the variance estimator $\hat{\sigma}^2$.
When estimating the probability for Joaquin to reach Category 5 status,
the eight nonzero family weights $W^{\left(i\right)}$
are 0.2, 0.2, 0.3, 0.3, 0.5, 0.6, 0.9, and 3.1.
The largest family weight of $W^{\left(i\right)} = 3.1$ accounts 
for $85\%$ of the variance $\hat{\sigma}^2$.
Without this particular $W^{\left(i\right)}$ value,
$\hat{\sigma}^2$ would be 3.5 times smaller for Quantile DMC compared to direct sampling. 
Because of this particular $W^{\left(i\right)}$ value,
$\hat{\sigma}^2$ is instead 1.9 times larger for Quantile DMC compared to direct sampling.

The largest weight of $W^{\left(i\right)} = 3.1$ belongs to a family
of late-developing storms.
The initial simulation in this family exhibited low intensity
at the first resampling time,
but it was not killed during the resampling step due to random chance.
Over time, the simulation increased in intensity.
The single particle was eventually split
into six family members,
and ultimately all six family members achieved Category 5 status.
The high variance of Joaquin $P_{life}$ estimates
can be wholly attributed to this single family of late-developing storms.

The story of the largest family weight $W^{\left(i\right)} = 3.1$
illustrates key areas in which 
our design of Quantile DMC simulations of tropical cyclones
was not optimal.
First, the reaction coordinate $\theta$ failed to anticipate 
which trajectories would lead to high intensities at later times.
At the first resampling time, the value of $\theta\left(\xi_1^{\left(i\right)}\right)$ 
was low,
but all six descendents of $\xi_1^{\left(i\right)}$
would go on to achieve Category 5 status.
This rapid intensification was not predictable
using sea surface pressure as our reaction coordinate.
However, an improved reaction coordinate $\theta$ could predict
lifetime intensity based on additional variables such
as steering flow, vertical wind shear, and relative humidity.
With an improved reaction coordinate,
future intensity could be more accurately identified,
thereby reducing Quantile DMC variance.

A second shortcoming in the design of simulations
was the poor identification of the time $T$ for Joaquin to reach peak intensity.
For Earl, the reference time $T$ was correctly identified as September 1 00:00 UTC,
the approximate time when storms achieved peak intensity in direct sampling
and Quantile DMC runs.
For Joaquin, on the other hand, the reference time $T$ was incorrectly identified as October 3 00:00 UTC,
nearly twenty-four hours before peak intensity occurred in direct sampling
and Quantile DMC runs.
With a later reference time $T$,
late-developing storms could be more appropriately sampled,
reducing the variance of $P_{life}$ estimates for Hurricane Joaquin.

We report three takeaway messages for future applications of Quantile DMC
to study intense tropical cyclones.

First, with an increased ensemble of $N = 1000$ simulations,
we are optimistic that Quantile DMC can provide low-error
estimates of extreme tail probabilities.
Already with a small sample of $N = 100$ simulations,
we see signs of reduced variance using Quantile DMC.
But with increased sample size $N$, 
we expect the error of Quantile DMC to shrink
with a faster-than-$N^{-1 / 2}$ scaling,
enhancing Quantile DMC's advantages over direct sampling.
When estimating the most extreme quantiles of intensity,
a splitting method such as Quantile DMC truly excels.

Second, when simulating cyclones, Quantile DMC 
is a more convenient algorithm to use than standard DMC.
Using Quantile DMC, the decision
to select a splitting constant of $C = 1$ is straightforward.
In contrast, using DMC, similar results can only be obtained with the foresight
to select a splitting constant $C$ with $C^{-1}$ being the standard
deviation of intensity, namely 7hPa to 8hPa.
Moreover, in future simulations of tropical cyclones,
there is the possibility of highly skewed or bimodal intensity distributions \cite{lee2016rapid}.
In these contexts, the additional robustness of Quantile DMC over DMC
may provide a further benefit.

Lastly, to maximize the potential benefit of Quantile DMC in future simulations,
it is of paramount importance to improve the reaction coordinate $\theta$ and 
estimates of the time of maximum intensity .
In our simulations, the poor identification of these parameters
led to added variance in some Quantile DMC calculations.
Improved parameters $\theta$ and $T$ would better predict lifetime storm intensities,
alleviating the problem of variance inflation.


The search for improved predictions of lifetime intensity is challenging in part
due to a limited understanding of the precursors and dynamics of rapid intensification \cite{montgomery2014paradigms}.
In complementary work, we develop a rare event analysis tool
that offers insight into the physics of TC rapid intensification \cite{plotkin2018maximizing}.
Such a technique could potentially identify a more suitable and predictive reaction coordinate
for future Quantile DMC applications.
By incorporating careful data analysis
and scientific insight into the model being simulated,
future work can potentially improve the efficiency of Quantile DMC
in tropical cyclone simulations.




\section{Conclusion}{\label{sec:Conclusion}}

Efficient sampling of extreme mesoscale weather remains one of the outstanding
computational challenges of the 21st century.
Extreme weather, such as tropical cyclones, squall lines, and floods, has a tremendous impact on human society,
yet assessing the frequency of extreme weather in past, current, and projected future
climates is extremely difficult.
Responding to this challenge,
we have introduced a new rare event sampling algorithm, Quantile DMC.
Combining Quantile DMC with dynamical downscaling
provides a new paradigm for calculating extreme weather statistics.
This approach potentially enables
high-accuracy, computationally-intensive models to be run with
reduced computational cost,
raising the quality of extreme weather statistics.

In Sections \ref{sub:QDMC} and \ref{sub:implementation},
we have provided a practical guide to using Quantile DMC.
In particular, we offer specific recommendations for the parameters 
to be used in the algorithm.
When computing tail probabilities for the Ornstein-Uhlenbeck process,
Quantile DMC is over a thousand times more efficient than
direct sampling and is more stable than Diffusion Monte Carlo.
When computing tail probabilities for intense tropical cyclones,
Quantile DMC is two to ten times more efficient than direct sampling,
with the possibility for greater efficiency in future simulations.

There remain important challenges in applying 
Quantile DMC to simulate extreme weather events.
In our simulations of tropical cyclones, we observe that
Quantile DMC's performance depends on
a \emph{reaction coordinate}, 
a one-dimensional coordinate that anticipates the occurrence of 
high-intensity weather.
The reaction coordinate that we used in our simulations was not optimal,
and we anticipate using an improved reaction coordinate in future TC sampling.
Fortunately, even with an imperfect choice of reaction coordinate,
it is possible for Quantile DMC to provide a reduction of variance.

We acknowledge two issues
that affect the future of rare event sampling of extreme weather.
First, splitting methods like Quantile DMC can only be successful if 
extreme weather is simulated stochastically.
However, stochastic models are easily available and increasingly used in many 
state-of-the-art geophysical computations \cite{berner2017stochastic},
so this does not present a major limitation in practice.
Second, rare event sampling is unnecessary if there exist models that
are both accurate and inexpensive to run.
While we acknowledge some statistical models and simplified physics models perform remarkably well
in today's extreme weather calculations \cite{emanuel2008hurricanes, maraun2010precipitation},
our outlook is to the future.
We believe high-resolution three-dimensional
models will eventually provide the greatest accuracy in all areas of extreme weather inference.
Developing rare event sampling methods is therefore
essential preparation for the future
as computationally intensive ensembles become the authoritative 
source of insight in extreme weather inference.


\begin{acknowledgments}

We acknowledge support from that National Science Foundation
under NSF award number 1623064.
RW and JW are supported by the Advanced Scientific Computing
Research Program within the DOE Office of Science through award DE-SC0014205.
RW was supported by NSF RTG award number 1547396
at the University of Chicago and by a MacCracken Fellowship at New York University.
DP was supported by the Department of Energy Computational Science
Graduate Fellowship Program of the Office of Science and National Nuclear Security.
MON was partially supported by the T.C. Chamberlin Postdoctoral
Fellowship at the University of Chicago.
The work was completed with resources provided by the University
of Chicago Research Computing Center.
RW acknowledges Alicia Zhao for her gracious and patient editorial assistance.

\end{acknowledgments}

\section{Appendix}

\subsection{Resampling schemes}

There are many possible \emph{resampling schemes} that can be used during DMC's resampling step.
One scheme that works well in practice is \emph{sorted stratified resampling} \cite{kitagawa1996monte, webber2019resampling}.
The scheme first sorts particles $\left(\xi_k^{\left(i\right)}\right)_{1 \leq i \leq N}$
based on the values of $\left(\theta\left(\xi_k^{\left(i\right)}\right)\right)_{1 \leq i \leq N}$
and then selects new particles $\left(\hat{\xi}_k^{\left(j\right)}\right)_{1 \leq j \leq N}$
using stratified resampling.

\begin{definition}[Sorted stratified resampling]{\label{def:sorted}}
~
\begin{enumerate}
\item
Sorting: Reindex the particles and weights  $\left(w_k^{\left(i\right)}, \xi_k^{\left(i\right)}\right)_{1 \leq i \leq N}$ so that
\begin{equation}
\theta\left(\xi_k^{\left(1\right)}\right) \leq \theta\left(\xi_k^{\left(2\right)}\right)
\leq \cdots \leq \theta\left(\xi_k^{\left(N\right)}\right)
\end{equation}
\item
Stratified resampling:
Construct the empirical quantile function
$Q_t \colon \left[0, 1\right) \rightarrow \mathbb{R}^d$ for the ensemble
$\left(w_k^{\left(i\right)}, \xi_k^{\left(i\right)}\right)_{1 \leq i \leq N}$
as follows:
\begin{equation}
Q_k\left(x\right) = \xi_k^{\left(i\right)}, 
\quad \frac{\sum_{j=1}^{i-1} w_k^{\left(j\right)}}{\sum_{j=1}^{N} w_k^{\left(j\right)}}
\leq x < \frac{\sum_{j=1}^i w_k^{\left(j\right)}}{\sum_{j=1}^{N} w_k^{\left(j\right)}}
\end{equation}
Select updated particles
$\hat{\xi}_k^{\left(j\right)} = Q_k\left(\frac{j - 1 + U_k^{\left(j\right)}}{N}\right)$
for $1 \leq j \leq N$,
where $U_k^{\left(j\right)}$ are independent $\Unif\left(0, 1\right)$ random variables.
\end{enumerate}
\end{definition}

It can be checked that each
particle $\xi_k^{\left(i\right)}$ is duplicated an expected number of
$w_k^{\left(i\right)} \slash \overline{w}_k$ times.
Therefore, sorted stratified resampling is a valid resampling scheme.

\subsection{Invariance under monotonic transformations}

Quantile DMC is unchanged if the reaction coordinate $\theta$
is replaced with a reaction coordinate $\tilde{\theta}$
that is a monotonic, one-to-one transformation of $\theta$.
To show this we provide an alternate description of Quantile DMC's 
process for splitting/killing particles that makes clear the property of invariance under monotonic transformations.

First, we introduce an order relation $x \prec y$ which indicates $\theta\left(x\right) < \theta\left(y\right)$
and an equivalence relation $x \sim y$ which indicates  $\theta\left(x\right) = \theta\left(y\right)$.
We observe that the order relation $x \prec y$ and equivalence relation $x \sim y$
remain unchanged if $\theta$ is replaced by $\tilde{\theta}$.

Second, define the quantiles $p_k^{\left(i\right)}$ with the formula
\begin{equation}
\sum_{j=1}^N z_k^{\left(j\right)}
\left[\mathds{1}\left\{\xi_k^{\left(j\right)} \prec \xi_k^{\left(i\right)}\right\} 
+ \frac{1}{2} \mathds{1}\left\{\xi_k^{\left(j\right)} \sim \xi_k^{\left(i\right)}\right\}\right]
\end{equation}
where 
\begin{equation}
\begin{cases}
z_k^{\left(i\right)} = \frac{1}{N}, 
& k = 0 \\
z_k^{\left(i\right)} = 
\frac{\exp\left\{-V_{k-1}^{\prime}\left(\hat{\xi}_{k-1}^{\left(i\right)}\right)\right\}}
{\sum_{j = 1}^N \exp\left\{-V_{k-1}^{\prime}\left(\hat{\xi}_{k-1}^{\left(j\right)}\right)\right\}},
& k > 0
\end{cases}
\end{equation}
The quantitities  $p_k^{\left(i\right)}$ are approximate quantiles for the distribution $\theta\left(X_{t_k}\right)$.
Thus, if $p_k^{\left(i\right)} = .9$ there is an approximate one-in-ten chance that 
$\theta\left(X_{t_k}\right)$ takes a value as high as $\theta\left(\xi_k^{\left(i\right)}\right)$.

Third, the splitting function takes the values
\begin{equation}
V_k^{\prime}\left(\xi_k^{\left(i\right)}\right) = V_k^{\prime}\left(F_{\nu_k}^{-1} \left(p_k^{\left(i\right)}\right)\right)
\end{equation}

In this description Quantile DMC only relies on the reaction coordinate $\theta$ through the order relation
$x \prec y$ and equivalence relation $x \sim y$.
These two relations are unchanged under monotonic, one-to-one transformations of the reaction coordinate.

In addition to the property of invariance under monotonic transformations,
Quantile DMC also has the property that estimates are unbiased,
which can be established following standard martingale arguments 
\cite{del2004feynman, webber2019resampling}.
Numerical evidence indicates that estimates converge with an asymptotic $1 \slash \sqrt{N}$ error rate,
as $N \rightarrow \infty$.
However, a rigorous mathematical analysis of Quantile DMC's error remains a task for future research.

\subsection{Variance estimation for Quantile DMC}

The variance estimator \eqref{advanced} for Quantile DMC was originally
developed assuming a different resampling scheme
called Bernoulli resampling is used \cite{chan2013general}.
Indeed, when the DMC algorithm is performed using Bernoulli resampling,
the variance estimator is asympotically consistent as $N \rightarrow \infty$.
We have chosen to use a different resampling scheme which gives better performance
than the Bernoulli resampling scheme.
Consequently the variance estimator $\hat{\sigma}^2$ is biased toward overestimating
the variance.
Numerical experiments with the Ornstein-Uhlenbeck process
suggest that the variance is inflated by less than $10\%$, at least for
$K \leq 10$ resampling times.
Another potential concern is the natural variability in the estimator 
$\hat{\sigma}^2$. 
We find that $\hat{\sigma}^2$ is most reliable when family weights
\begin{equation}
W^{\left(i\right)} = \sum_{\anc\left(\xi_k^{\left(j\right)}\right) = i}
\frac{\overline{w}_{K-1} \mathds{1}\left\{\xi_K^{\left(j\right)} \in A\right\}}
{\exp\left\{V_{K-1}^{\prime}\left(\hat{\xi}_{K-1}^{\left(j\right)}\right)\right\}}
\end{equation}
are nonzero for many indices $i$.
Consequently, $\hat{\sigma}^2$ is most reliable 
when sample size $N$ is high
and the number of resampling times $K$ is low.
We find that the variance estimator $\hat{\sigma}^2$ provides a
valuable tool for gauging the quality of Quantile DMC estimates.
We caution the reader, however, there may be situations outside the current
context where the variance estimator $\hat{\sigma}^2$ behaves poorly.

\bibliography{arxiv}

\begin{thebibliography}{53}%
\makeatletter
\providecommand \@ifxundefined [1]{%
 \@ifx{#1\undefined}
}%
\providecommand \@ifnum [1]{%
 \ifnum #1\expandafter \@firstoftwo
 \else \expandafter \@secondoftwo
 \fi
}%
\providecommand \@ifx [1]{%
 \ifx #1\expandafter \@firstoftwo
 \else \expandafter \@secondoftwo
 \fi
}%
\providecommand \natexlab [1]{#1}%
\providecommand \enquote  [1]{``#1''}%
\providecommand \bibnamefont  [1]{#1}%
\providecommand \bibfnamefont [1]{#1}%
\providecommand \citenamefont [1]{#1}%
\providecommand \href@noop [0]{\@secondoftwo}%
\providecommand \href [0]{\begingroup \@sanitize@url \@href}%
\providecommand \@href[1]{\@@startlink{#1}\@@href}%
\providecommand \@@href[1]{\endgroup#1\@@endlink}%
\providecommand \@sanitize@url [0]{\catcode `\\12\catcode `\$12\catcode
  `\&12\catcode `\#12\catcode `\^12\catcode `\_12\catcode `\%12\relax}%
\providecommand \@@startlink[1]{}%
\providecommand \@@endlink[0]{}%
\providecommand \url  [0]{\begingroup\@sanitize@url \@url }%
\providecommand \@url [1]{\endgroup\@href {#1}{\urlprefix }}%
\providecommand \urlprefix  [0]{URL }%
\providecommand \Eprint [0]{\href }%
\providecommand \doibase [0]{http://dx.doi.org/}%
\providecommand \selectlanguage [0]{\@gobble}%
\providecommand \bibinfo  [0]{\@secondoftwo}%
\providecommand \bibfield  [0]{\@secondoftwo}%
\providecommand \translation [1]{[#1]}%
\providecommand \BibitemOpen [0]{}%
\providecommand \bibitemStop [0]{}%
\providecommand \bibitemNoStop [0]{.\EOS\space}%
\providecommand \EOS [0]{\spacefactor3000\relax}%
\providecommand \BibitemShut  [1]{\csname bibitem#1\endcsname}%
\let\auto@bib@innerbib\@empty
\bibitem [{\citenamefont {Meehl}\ and\ \citenamefont
  {Tebaldi}(2004)}]{meehl2004more}%
  \BibitemOpen
  \bibfield  {author} {\bibinfo {author} {\bibfnamefont {G.~A.}\ \bibnamefont
  {Meehl}}\ and\ \bibinfo {author} {\bibfnamefont {C.}~\bibnamefont
  {Tebaldi}},\ }\bibfield  {title} {\enquote {\bibinfo {title} {More intense,
  more frequent, and longer lasting heat waves in the 21st century},}\
  }\href@noop {} {\bibfield  {journal} {\bibinfo  {journal} {Science}\ }\textbf
  {\bibinfo {volume} {305}},\ \bibinfo {pages} {994--997} (\bibinfo {year}
  {2004})}\BibitemShut {NoStop}%
\bibitem [{\citenamefont {Bender}\ \emph {et~al.}(2010)\citenamefont {Bender},
  \citenamefont {Knutson}, \citenamefont {Tuleya}, \citenamefont {Sirutis},
  \citenamefont {Vecchi}, \citenamefont {Garner},\ and\ \citenamefont
  {Held}}]{bender2010modeled}%
  \BibitemOpen
  \bibfield  {author} {\bibinfo {author} {\bibfnamefont {M.~A.}\ \bibnamefont
  {Bender}}, \bibinfo {author} {\bibfnamefont {T.~R.}\ \bibnamefont {Knutson}},
  \bibinfo {author} {\bibfnamefont {R.~E.}\ \bibnamefont {Tuleya}}, \bibinfo
  {author} {\bibfnamefont {J.~J.}\ \bibnamefont {Sirutis}}, \bibinfo {author}
  {\bibfnamefont {G.~A.}\ \bibnamefont {Vecchi}}, \bibinfo {author}
  {\bibfnamefont {S.~T.}\ \bibnamefont {Garner}}, \ and\ \bibinfo {author}
  {\bibfnamefont {I.~M.}\ \bibnamefont {Held}},\ }\bibfield  {title} {\enquote
  {\bibinfo {title} {Modeled impact of anthropogenic warming on the frequency
  of intense {A}tlantic hurricanes},}\ }\href@noop {} {\bibfield  {journal}
  {\bibinfo  {journal} {Science}\ }\textbf {\bibinfo {volume} {327}},\ \bibinfo
  {pages} {454--458} (\bibinfo {year} {2010})}\BibitemShut {NoStop}%
\bibitem [{\citenamefont {Lin}\ and\ \citenamefont
  {Emanuel}(2016)}]{lin2016grey}%
  \BibitemOpen
  \bibfield  {author} {\bibinfo {author} {\bibfnamefont {N.}~\bibnamefont
  {Lin}}\ and\ \bibinfo {author} {\bibfnamefont {K.}~\bibnamefont {Emanuel}},\
  }\bibfield  {title} {\enquote {\bibinfo {title} {Grey swan tropical
  cyclones},}\ }\href@noop {} {\bibfield  {journal} {\bibinfo  {journal}
  {Nature Climate Change}\ }\textbf {\bibinfo {volume} {6}},\ \bibinfo {pages}
  {106--111} (\bibinfo {year} {2016})}\BibitemShut {NoStop}%
\bibitem [{\citenamefont {Kahn}\ and\ \citenamefont
  {Harris}(1951)}]{kahn1951estimation}%
  \BibitemOpen
  \bibfield  {author} {\bibinfo {author} {\bibfnamefont {H.}~\bibnamefont
  {Kahn}}\ and\ \bibinfo {author} {\bibfnamefont {T.~E.}\ \bibnamefont
  {Harris}},\ }\bibfield  {title} {\enquote {\bibinfo {title} {Estimation of
  particle transmission by random sampling},}\ }\href@noop {} {\bibfield
  {journal} {\bibinfo  {journal} {National Bureau of Standards Applied
  Mathematics Series}\ }\textbf {\bibinfo {volume} {12}},\ \bibinfo {pages}
  {27--30} (\bibinfo {year} {1951})}\BibitemShut {NoStop}%
\bibitem [{\citenamefont {Rosenbluth}\ and\ \citenamefont
  {Rosenbluth}(1955)}]{rosenbluth1955monte}%
  \BibitemOpen
  \bibfield  {author} {\bibinfo {author} {\bibfnamefont {M.~N.}\ \bibnamefont
  {Rosenbluth}}\ and\ \bibinfo {author} {\bibfnamefont {A.~W.}\ \bibnamefont
  {Rosenbluth}},\ }\bibfield  {title} {\enquote {\bibinfo {title} {Monte
  {C}arlo calculation of the average extension of molecular chains},}\
  }\href@noop {} {\bibfield  {journal} {\bibinfo  {journal} {The Journal of
  Chemical Physics}\ }\textbf {\bibinfo {volume} {23}},\ \bibinfo {pages}
  {356--359} (\bibinfo {year} {1955})}\BibitemShut {NoStop}%
\bibitem [{\citenamefont {Dubi}, \citenamefont {Elperin},\ and\ \citenamefont
  {Dudziak}(1982)}]{dubi1982geometrical}%
  \BibitemOpen
  \bibfield  {author} {\bibinfo {author} {\bibfnamefont {A.}~\bibnamefont
  {Dubi}}, \bibinfo {author} {\bibfnamefont {T.}~\bibnamefont {Elperin}}, \
  and\ \bibinfo {author} {\bibfnamefont {D.~J.}\ \bibnamefont {Dudziak}},\
  }\bibfield  {title} {\enquote {\bibinfo {title} {Geometrical splitting in
  {M}onte {C}arlo},}\ }\href@noop {} {\bibfield  {journal} {\bibinfo  {journal}
  {Nuclear Science and Engineering}\ }\textbf {\bibinfo {volume} {80}},\
  \bibinfo {pages} {139--161} (\bibinfo {year} {1982})}\BibitemShut {NoStop}%
\bibitem [{\citenamefont {Huber}\ and\ \citenamefont
  {Kim}(1996)}]{huber1996weighted}%
  \BibitemOpen
  \bibfield  {author} {\bibinfo {author} {\bibfnamefont {G.~A.}\ \bibnamefont
  {Huber}}\ and\ \bibinfo {author} {\bibfnamefont {S.}~\bibnamefont {Kim}},\
  }\bibfield  {title} {\enquote {\bibinfo {title} {Weighted-ensemble {B}rownian
  dynamics simulations for protein association reactions.}}\ }\href@noop {}
  {\bibfield  {journal} {\bibinfo  {journal} {Biophysical Journal}\ }\textbf
  {\bibinfo {volume} {70}},\ \bibinfo {pages} {97} (\bibinfo {year}
  {1996})}\BibitemShut {NoStop}%
\bibitem [{\citenamefont {Glasserman}, \citenamefont {Heidelberger},\ and\
  \citenamefont {Shahabuddin}(2000)}]{glasserman2000variance}%
  \BibitemOpen
  \bibfield  {author} {\bibinfo {author} {\bibfnamefont {P.}~\bibnamefont
  {Glasserman}}, \bibinfo {author} {\bibfnamefont {P.}~\bibnamefont
  {Heidelberger}}, \ and\ \bibinfo {author} {\bibfnamefont {P.}~\bibnamefont
  {Shahabuddin}},\ }\bibfield  {title} {\enquote {\bibinfo {title} {Variance
  reduction techniques for estimating value-at-risk},}\ }\href@noop {}
  {\bibfield  {journal} {\bibinfo  {journal} {Management Science}\ }\textbf
  {\bibinfo {volume} {46}},\ \bibinfo {pages} {1349--1364} (\bibinfo {year}
  {2000})}\BibitemShut {NoStop}%
\bibitem [{\citenamefont {Hoffman}\ \emph {et~al.}(2006)\citenamefont
  {Hoffman}, \citenamefont {Henderson}, \citenamefont {Leidner}, \citenamefont
  {Grassotti},\ and\ \citenamefont {Nehrkorn}}]{hoffman2006response}%
  \BibitemOpen
  \bibfield  {author} {\bibinfo {author} {\bibfnamefont {R.~N.}\ \bibnamefont
  {Hoffman}}, \bibinfo {author} {\bibfnamefont {J.~M.}\ \bibnamefont
  {Henderson}}, \bibinfo {author} {\bibfnamefont {S.~M.}\ \bibnamefont
  {Leidner}}, \bibinfo {author} {\bibfnamefont {C.}~\bibnamefont {Grassotti}},
  \ and\ \bibinfo {author} {\bibfnamefont {T.}~\bibnamefont {Nehrkorn}},\
  }\bibfield  {title} {\enquote {\bibinfo {title} {The response of damaging
  winds of a simulated tropical cyclone to finite-amplitude perturbations of
  different variables},}\ }\href@noop {} {\bibfield  {journal} {\bibinfo
  {journal} {Journal of the Atmospheric Sciences}\ }\textbf {\bibinfo {volume}
  {63}},\ \bibinfo {pages} {1924--1937} (\bibinfo {year} {2006})}\BibitemShut
  {NoStop}%
\bibitem [{\citenamefont {Weare}(2009)}]{weare2009particle}%
  \BibitemOpen
  \bibfield  {author} {\bibinfo {author} {\bibfnamefont {J.}~\bibnamefont
  {Weare}},\ }\bibfield  {title} {\enquote {\bibinfo {title} {Particle
  filtering with path sampling and an application to a bimodal ocean current
  model},}\ }\href@noop {} {\bibfield  {journal} {\bibinfo  {journal} {Journal
  of Computational Physics}\ }\textbf {\bibinfo {volume} {228}},\ \bibinfo
  {pages} {4312--4331} (\bibinfo {year} {2009})}\BibitemShut {NoStop}%
\bibitem [{\citenamefont {Vanden-Eijnden}\ and\ \citenamefont
  {Weare}(2013)}]{vanden2013data}%
  \BibitemOpen
  \bibfield  {author} {\bibinfo {author} {\bibfnamefont {E.}~\bibnamefont
  {Vanden-Eijnden}}\ and\ \bibinfo {author} {\bibfnamefont {J.}~\bibnamefont
  {Weare}},\ }\bibfield  {title} {\enquote {\bibinfo {title} {Data assimilation
  in the low noise regime with application to the {K}uroshio},}\ }\href@noop {}
  {\bibfield  {journal} {\bibinfo  {journal} {Monthly Weather Review}\ }\textbf
  {\bibinfo {volume} {141}},\ \bibinfo {pages} {1822--1841} (\bibinfo {year}
  {2013})}\BibitemShut {NoStop}%
\bibitem [{\citenamefont {Ragone}, \citenamefont {Wouters},\ and\ \citenamefont
  {Bouchet}(2018)}]{ragone2017computation}%
  \BibitemOpen
  \bibfield  {author} {\bibinfo {author} {\bibfnamefont {F.}~\bibnamefont
  {Ragone}}, \bibinfo {author} {\bibfnamefont {J.}~\bibnamefont {Wouters}}, \
  and\ \bibinfo {author} {\bibfnamefont {F.}~\bibnamefont {Bouchet}},\
  }\bibfield  {title} {\enquote {\bibinfo {title} {Computation of extreme heat
  waves in climate models using a large deviation algorithm},}\ }\href@noop {}
  {\bibfield  {journal} {\bibinfo  {journal} {Proceedings of the National
  Academy of Sciences}\ }\textbf {\bibinfo {volume} {115}},\ \bibinfo {pages}
  {24--29} (\bibinfo {year} {2018})}\BibitemShut {NoStop}%
\bibitem [{\citenamefont {Dematteis}, \citenamefont {Grafke},\ and\
  \citenamefont {Vanden-Eijnden}(2018)}]{dematteis2018rogue}%
  \BibitemOpen
  \bibfield  {author} {\bibinfo {author} {\bibfnamefont {G.}~\bibnamefont
  {Dematteis}}, \bibinfo {author} {\bibfnamefont {T.}~\bibnamefont {Grafke}}, \
  and\ \bibinfo {author} {\bibfnamefont {E.}~\bibnamefont {Vanden-Eijnden}},\
  }\bibfield  {title} {\enquote {\bibinfo {title} {Rogue waves and large
  deviations in deep sea},}\ }\href@noop {} {\bibfield  {journal} {\bibinfo
  {journal} {Proceedings of the National Academy of Sciences}\ }\textbf
  {\bibinfo {volume} {115}},\ \bibinfo {pages} {855--860} (\bibinfo {year}
  {2018})}\BibitemShut {NoStop}%
\bibitem [{\citenamefont {Plotkin}\ \emph {et~al.}(2019)\citenamefont
  {Plotkin}, \citenamefont {Webber}, \citenamefont {O'Neill}, \citenamefont
  {Weare},\ and\ \citenamefont {Abbot}}]{plotkin2018maximizing}%
  \BibitemOpen
  \bibfield  {author} {\bibinfo {author} {\bibfnamefont {D.~A.}\ \bibnamefont
  {Plotkin}}, \bibinfo {author} {\bibfnamefont {R.~J.}\ \bibnamefont {Webber}},
  \bibinfo {author} {\bibfnamefont {M.~E.}\ \bibnamefont {O'Neill}}, \bibinfo
  {author} {\bibfnamefont {J.}~\bibnamefont {Weare}}, \ and\ \bibinfo {author}
  {\bibfnamefont {D.~S.}\ \bibnamefont {Abbot}},\ }\bibfield  {title} {\enquote
  {\bibinfo {title} {Maximizing simulated tropical cyclone intensity with
  action minimization},}\ }\href@noop {} {\bibfield  {journal} {\bibinfo
  {journal} {Journal of Advances in Modeling Earth Systems}\ } (\bibinfo {year}
  {2019})}\BibitemShut {NoStop}%
\bibitem [{\citenamefont {Pielke~Jr}\ and\ \citenamefont
  {Downton}(2000)}]{pielke2000precipitation}%
  \BibitemOpen
  \bibfield  {author} {\bibinfo {author} {\bibfnamefont {R.~A.}\ \bibnamefont
  {Pielke~Jr}}\ and\ \bibinfo {author} {\bibfnamefont {M.~W.}\ \bibnamefont
  {Downton}},\ }\bibfield  {title} {\enquote {\bibinfo {title} {Precipitation
  and damaging floods: Trends in the {U}nited {S}tates, 1932--97},}\
  }\href@noop {} {\bibfield  {journal} {\bibinfo  {journal} {Journal of
  Climate}\ }\textbf {\bibinfo {volume} {13}},\ \bibinfo {pages} {3625--3637}
  (\bibinfo {year} {2000})}\BibitemShut {NoStop}%
\bibitem [{\citenamefont {Pielke~Jr}\ \emph {et~al.}(2008)\citenamefont
  {Pielke~Jr}, \citenamefont {Gratz}, \citenamefont {Landsea}, \citenamefont
  {Collins}, \citenamefont {Saunders},\ and\ \citenamefont
  {Musulin}}]{pielke2008normalized}%
  \BibitemOpen
  \bibfield  {author} {\bibinfo {author} {\bibfnamefont {R.~A.}\ \bibnamefont
  {Pielke~Jr}}, \bibinfo {author} {\bibfnamefont {J.}~\bibnamefont {Gratz}},
  \bibinfo {author} {\bibfnamefont {C.~W.}\ \bibnamefont {Landsea}}, \bibinfo
  {author} {\bibfnamefont {D.}~\bibnamefont {Collins}}, \bibinfo {author}
  {\bibfnamefont {M.~A.}\ \bibnamefont {Saunders}}, \ and\ \bibinfo {author}
  {\bibfnamefont {R.}~\bibnamefont {Musulin}},\ }\bibfield  {title} {\enquote
  {\bibinfo {title} {Normalized hurricane damage in the {U}nited {S}tates:
  1900--2005},}\ }\href@noop {} {\bibfield  {journal} {\bibinfo  {journal}
  {Natural Hazards Review}\ }\textbf {\bibinfo {volume} {9}},\ \bibinfo {pages}
  {29--42} (\bibinfo {year} {2008})}\BibitemShut {NoStop}%
\bibitem [{\citenamefont {Davis}\ \emph {et~al.}(2008)\citenamefont {Davis},
  \citenamefont {Wang}, \citenamefont {Chen}, \citenamefont {Chen},
  \citenamefont {Corbosiero}, \citenamefont {DeMaria}, \citenamefont {Dudhia},
  \citenamefont {Holland}, \citenamefont {Klemp}, \citenamefont {Michalakes}
  \emph {et~al.}}]{davis2008prediction}%
  \BibitemOpen
  \bibfield  {author} {\bibinfo {author} {\bibfnamefont {C.}~\bibnamefont
  {Davis}}, \bibinfo {author} {\bibfnamefont {W.}~\bibnamefont {Wang}},
  \bibinfo {author} {\bibfnamefont {S.~S.}\ \bibnamefont {Chen}}, \bibinfo
  {author} {\bibfnamefont {Y.}~\bibnamefont {Chen}}, \bibinfo {author}
  {\bibfnamefont {K.}~\bibnamefont {Corbosiero}}, \bibinfo {author}
  {\bibfnamefont {M.}~\bibnamefont {DeMaria}}, \bibinfo {author} {\bibfnamefont
  {J.}~\bibnamefont {Dudhia}}, \bibinfo {author} {\bibfnamefont
  {G.}~\bibnamefont {Holland}}, \bibinfo {author} {\bibfnamefont
  {J.}~\bibnamefont {Klemp}}, \bibinfo {author} {\bibfnamefont
  {J.}~\bibnamefont {Michalakes}},  \emph {et~al.},\ }\bibfield  {title}
  {\enquote {\bibinfo {title} {Prediction of landfalling hurricanes with the
  {A}dvanced {H}urricane {WRF} model},}\ }\href@noop {} {\bibfield  {journal}
  {\bibinfo  {journal} {Monthly Weather Review}\ }\textbf {\bibinfo {volume}
  {136}},\ \bibinfo {pages} {1990--2005} (\bibinfo {year} {2008})}\BibitemShut
  {NoStop}%
\bibitem [{\citenamefont {Hirabayashi}\ \emph {et~al.}(2008)\citenamefont
  {Hirabayashi}, \citenamefont {Kanae}, \citenamefont {Emori}, \citenamefont
  {Oki},\ and\ \citenamefont {Kimoto}}]{hirabayashi2008global}%
  \BibitemOpen
  \bibfield  {author} {\bibinfo {author} {\bibfnamefont {Y.}~\bibnamefont
  {Hirabayashi}}, \bibinfo {author} {\bibfnamefont {S.}~\bibnamefont {Kanae}},
  \bibinfo {author} {\bibfnamefont {S.}~\bibnamefont {Emori}}, \bibinfo
  {author} {\bibfnamefont {T.}~\bibnamefont {Oki}}, \ and\ \bibinfo {author}
  {\bibfnamefont {M.}~\bibnamefont {Kimoto}},\ }\bibfield  {title} {\enquote
  {\bibinfo {title} {Global projections of changing risks of floods and
  droughts in a changing climate},}\ }\href@noop {} {\bibfield  {journal}
  {\bibinfo  {journal} {Hydrological Sciences Journal}\ }\textbf {\bibinfo
  {volume} {53}},\ \bibinfo {pages} {754--772} (\bibinfo {year}
  {2008})}\BibitemShut {NoStop}%
\bibitem [{\citenamefont {Kalos}(1962)}]{kalos1962monte}%
  \BibitemOpen
  \bibfield  {author} {\bibinfo {author} {\bibfnamefont {M.}~\bibnamefont
  {Kalos}},\ }\bibfield  {title} {\enquote {\bibinfo {title} {Monte {C}arlo
  calculations of the ground state of three- and four-body nuclei},}\
  }\href@noop {} {\bibfield  {journal} {\bibinfo  {journal} {Physical Review}\
  }\textbf {\bibinfo {volume} {128}},\ \bibinfo {pages} {1791--1795} (\bibinfo
  {year} {1962})}\BibitemShut {NoStop}%
\bibitem [{\citenamefont {Grimm}\ and\ \citenamefont
  {Storer}(1971)}]{grimm1971monte}%
  \BibitemOpen
  \bibfield  {author} {\bibinfo {author} {\bibfnamefont {R.}~\bibnamefont
  {Grimm}}\ and\ \bibinfo {author} {\bibfnamefont {R.}~\bibnamefont {Storer}},\
  }\bibfield  {title} {\enquote {\bibinfo {title} {Monte-{C}arlo solution of
  {S}chr{\"o}dinger's equation},}\ }\href@noop {} {\bibfield  {journal}
  {\bibinfo  {journal} {Journal of Computational Physics}\ }\textbf {\bibinfo
  {volume} {7}},\ \bibinfo {pages} {134--156} (\bibinfo {year}
  {1971})}\BibitemShut {NoStop}%
\bibitem [{\citenamefont {Giardina}, \citenamefont {Kurchan},\ and\
  \citenamefont {Peliti}(2006)}]{giardina2006direct}%
  \BibitemOpen
  \bibfield  {author} {\bibinfo {author} {\bibfnamefont {C.}~\bibnamefont
  {Giardina}}, \bibinfo {author} {\bibfnamefont {J.}~\bibnamefont {Kurchan}}, \
  and\ \bibinfo {author} {\bibfnamefont {L.}~\bibnamefont {Peliti}},\
  }\bibfield  {title} {\enquote {\bibinfo {title} {Direct evaluation of
  large-deviation functions},}\ }\href@noop {} {\bibfield  {journal} {\bibinfo
  {journal} {Physical Review Letters}\ }\textbf {\bibinfo {volume} {96}},\
  \bibinfo {pages} {1--4} (\bibinfo {year} {2006})}\BibitemShut {NoStop}%
\bibitem [{\citenamefont {Hairer}\ and\ \citenamefont
  {Weare}(2014)}]{hairer2014improved}%
  \BibitemOpen
  \bibfield  {author} {\bibinfo {author} {\bibfnamefont {M.}~\bibnamefont
  {Hairer}}\ and\ \bibinfo {author} {\bibfnamefont {J.}~\bibnamefont {Weare}},\
  }\bibfield  {title} {\enquote {\bibinfo {title} {Improved diffusion {M}onte
  {C}arlo},}\ }\href@noop {} {\bibfield  {journal} {\bibinfo  {journal}
  {Communications on Pure and Applied Mathematics}\ }\textbf {\bibinfo {volume}
  {67}},\ \bibinfo {pages} {1995--2021} (\bibinfo {year} {2014})}\BibitemShut
  {NoStop}%
\bibitem [{\citenamefont {C{\'e}rou}\ \emph {et~al.}(2006)\citenamefont
  {C{\'e}rou}, \citenamefont {Del~Moral}, \citenamefont {LeGland},\ and\
  \citenamefont {Lezaud}}]{cerou2006genetic}%
  \BibitemOpen
  \bibfield  {author} {\bibinfo {author} {\bibfnamefont {F.}~\bibnamefont
  {C{\'e}rou}}, \bibinfo {author} {\bibfnamefont {P.}~\bibnamefont
  {Del~Moral}}, \bibinfo {author} {\bibfnamefont {F.}~\bibnamefont {LeGland}},
  \ and\ \bibinfo {author} {\bibfnamefont {P.}~\bibnamefont {Lezaud}},\
  }\bibfield  {title} {\enquote {\bibinfo {title} {Genetic genealogical models
  in rare event analysis},}\ }\href@noop {} {\bibfield  {journal} {\bibinfo
  {journal} {Alea}\ }\textbf {\bibinfo {volume} {1}},\ \bibinfo {pages}
  {181--203} (\bibinfo {year} {2006})}\BibitemShut {NoStop}%
\bibitem [{\citenamefont {Del~Moral}\ and\ \citenamefont
  {Garnier}(2005)}]{del2005genealogical}%
  \BibitemOpen
  \bibfield  {author} {\bibinfo {author} {\bibfnamefont {P.}~\bibnamefont
  {Del~Moral}}\ and\ \bibinfo {author} {\bibfnamefont {J.}~\bibnamefont
  {Garnier}},\ }\bibfield  {title} {\enquote {\bibinfo {title} {Genealogical
  particle analysis of rare events},}\ }\href@noop {} {\bibfield  {journal}
  {\bibinfo  {journal} {The Annals of Applied Probability}\ }\textbf {\bibinfo
  {volume} {15}},\ \bibinfo {pages} {2496--2534} (\bibinfo {year}
  {2005})}\BibitemShut {NoStop}%
\bibitem [{\citenamefont {Wouters}\ and\ \citenamefont
  {Bouchet}(2016)}]{wouters2016rare}%
  \BibitemOpen
  \bibfield  {author} {\bibinfo {author} {\bibfnamefont {J.}~\bibnamefont
  {Wouters}}\ and\ \bibinfo {author} {\bibfnamefont {F.}~\bibnamefont
  {Bouchet}},\ }\bibfield  {title} {\enquote {\bibinfo {title} {Rare event
  computation in deterministic chaotic systems using genealogical particle
  analysis},}\ }\href@noop {} {\bibfield  {journal} {\bibinfo  {journal}
  {Journal of Physics A: Mathematical and Theoretical}\ }\textbf {\bibinfo
  {volume} {49}},\ \bibinfo {pages} {1--24} (\bibinfo {year}
  {2016})}\BibitemShut {NoStop}%
\bibitem [{\citenamefont {Del~Moral}(2004)}]{del2004feynman}%
  \BibitemOpen
  \bibfield  {author} {\bibinfo {author} {\bibfnamefont {P.}~\bibnamefont
  {Del~Moral}},\ }\href@noop {} {\emph {\bibinfo {title} {Feynman-{K}ac
  Formulae: Genealogical and Interacting Particle Systems with Applications}}}\
  (\bibinfo  {publisher} {Springer Science \& Business Media},\ \bibinfo {year}
  {2004})\BibitemShut {NoStop}%
\bibitem [{\citenamefont {Webber}(2019)}]{webber2019resampling}%
  \BibitemOpen
  \bibfield  {author} {\bibinfo {author} {\bibfnamefont {R.~J.}\ \bibnamefont
  {Webber}},\ }\bibfield  {title} {\enquote {\bibinfo {title} {Unifying
  {S}equential {M}onte {C}arlo with resampling matrices},}\ }\href@noop {}
  {\bibfield  {journal} {\bibinfo  {journal} {arXiv:1903.12583 [math.NA]}\ }
  (\bibinfo {year} {2019})}\BibitemShut {NoStop}%
\bibitem [{\citenamefont {Kitagawa}(1996)}]{kitagawa1996monte}%
  \BibitemOpen
  \bibfield  {author} {\bibinfo {author} {\bibfnamefont {G.}~\bibnamefont
  {Kitagawa}},\ }\bibfield  {title} {\enquote {\bibinfo {title} {Monte {C}arlo
  filter and smoother for non-{G}aussian nonlinear state space models},}\
  }\href@noop {} {\bibfield  {journal} {\bibinfo  {journal} {Journal of
  Computational and Graphical Statistics}\ }\textbf {\bibinfo {volume} {5}},\
  \bibinfo {pages} {1--25} (\bibinfo {year} {1996})}\BibitemShut {NoStop}%
\bibitem [{\citenamefont {Rachev}\ and\ \citenamefont
  {R{\"u}schendorf}(1998)}]{rachev1998mass}%
  \BibitemOpen
  \bibfield  {author} {\bibinfo {author} {\bibfnamefont {S.~T.}\ \bibnamefont
  {Rachev}}\ and\ \bibinfo {author} {\bibfnamefont {L.}~\bibnamefont
  {R{\"u}schendorf}},\ }\href@noop {} {\emph {\bibinfo {title} {Mass
  Transportation Problems: Volume I: Theory}}},\ Vol.~\bibinfo {volume} {1}\
  (\bibinfo  {publisher} {Springer Science \& Business Media},\ \bibinfo {year}
  {1998})\BibitemShut {NoStop}%
\bibitem [{\citenamefont {C{\'e}rou}\ and\ \citenamefont
  {Guyader}(2007)}]{cerou2007adaptive}%
  \BibitemOpen
  \bibfield  {author} {\bibinfo {author} {\bibfnamefont {F.}~\bibnamefont
  {C{\'e}rou}}\ and\ \bibinfo {author} {\bibfnamefont {A.}~\bibnamefont
  {Guyader}},\ }\bibfield  {title} {\enquote {\bibinfo {title} {Adaptive
  multilevel splitting for rare event analysis},}\ }\href@noop {} {\bibfield
  {journal} {\bibinfo  {journal} {Stochastic Analysis and Applications}\
  }\textbf {\bibinfo {volume} {25}},\ \bibinfo {pages} {417--443} (\bibinfo
  {year} {2007})}\BibitemShut {NoStop}%
\bibitem [{\citenamefont {Guttenberg}, \citenamefont {Dinner},\ and\
  \citenamefont {Weare}(2012)}]{guttenberg2012steered}%
  \BibitemOpen
  \bibfield  {author} {\bibinfo {author} {\bibfnamefont {N.}~\bibnamefont
  {Guttenberg}}, \bibinfo {author} {\bibfnamefont {A.~R.}\ \bibnamefont
  {Dinner}}, \ and\ \bibinfo {author} {\bibfnamefont {J.}~\bibnamefont
  {Weare}},\ }\bibfield  {title} {\enquote {\bibinfo {title} {Steered
  transition path sampling},}\ }\href@noop {} {\bibfield  {journal} {\bibinfo
  {journal} {The Journal of Chemical Physics}\ }\textbf {\bibinfo {volume}
  {136}},\ \bibinfo {pages} {1--11} (\bibinfo {year} {2012})}\BibitemShut
  {NoStop}%
\bibitem [{\citenamefont {Liu}(2008)}]{liu2008monte}%
  \BibitemOpen
  \bibfield  {author} {\bibinfo {author} {\bibfnamefont {J.~S.}\ \bibnamefont
  {Liu}},\ }\href@noop {} {\emph {\bibinfo {title} {Monte {C}arlo strategies in
  scientific computing}}}\ (\bibinfo  {publisher} {Springer Science \& Business
  Media},\ \bibinfo {year} {2008})\BibitemShut {NoStop}%
\bibitem [{\citenamefont {Chan}, \citenamefont {Lai}\ \emph
  {et~al.}(2013)\citenamefont {Chan}, \citenamefont {Lai} \emph
  {et~al.}}]{chan2013general}%
  \BibitemOpen
  \bibfield  {author} {\bibinfo {author} {\bibfnamefont {H.~P.}\ \bibnamefont
  {Chan}}, \bibinfo {author} {\bibfnamefont {T.~L.}\ \bibnamefont {Lai}},
  \emph {et~al.},\ }\bibfield  {title} {\enquote {\bibinfo {title} {A general
  theory of particle filters in hidden {M}arkov models and some
  applications},}\ }\href@noop {} {\bibfield  {journal} {\bibinfo  {journal}
  {The Annals of Statistics}\ }\textbf {\bibinfo {volume} {41}},\ \bibinfo
  {pages} {2877--2904} (\bibinfo {year} {2013})}\BibitemShut {NoStop}%
\bibitem [{\citenamefont {Coumou}\ and\ \citenamefont
  {Rahmstorf}(2012)}]{coumou2012decade}%
  \BibitemOpen
  \bibfield  {author} {\bibinfo {author} {\bibfnamefont {D.}~\bibnamefont
  {Coumou}}\ and\ \bibinfo {author} {\bibfnamefont {S.}~\bibnamefont
  {Rahmstorf}},\ }\bibfield  {title} {\enquote {\bibinfo {title} {A decade of
  weather extremes},}\ }\href@noop {} {\bibfield  {journal} {\bibinfo
  {journal} {Nature Climate Change}\ }\textbf {\bibinfo {volume} {2}},\
  \bibinfo {pages} {491--496} (\bibinfo {year} {2012})}\BibitemShut {NoStop}%
\bibitem [{\citenamefont {Knutson}\ \emph {et~al.}(2010)\citenamefont
  {Knutson}, \citenamefont {McBride}, \citenamefont {Chan}, \citenamefont
  {Emanuel}, \citenamefont {Holland}, \citenamefont {Landsea}, \citenamefont
  {Held}, \citenamefont {Kossin}, \citenamefont {Srivastava},\ and\
  \citenamefont {Sugi}}]{knutson2010tropical}%
  \BibitemOpen
  \bibfield  {author} {\bibinfo {author} {\bibfnamefont {T.~R.}\ \bibnamefont
  {Knutson}}, \bibinfo {author} {\bibfnamefont {J.~L.}\ \bibnamefont
  {McBride}}, \bibinfo {author} {\bibfnamefont {J.}~\bibnamefont {Chan}},
  \bibinfo {author} {\bibfnamefont {K.}~\bibnamefont {Emanuel}}, \bibinfo
  {author} {\bibfnamefont {G.}~\bibnamefont {Holland}}, \bibinfo {author}
  {\bibfnamefont {C.}~\bibnamefont {Landsea}}, \bibinfo {author} {\bibfnamefont
  {I.}~\bibnamefont {Held}}, \bibinfo {author} {\bibfnamefont {J.~P.}\
  \bibnamefont {Kossin}}, \bibinfo {author} {\bibfnamefont {A.}~\bibnamefont
  {Srivastava}}, \ and\ \bibinfo {author} {\bibfnamefont {M.}~\bibnamefont
  {Sugi}},\ }\bibfield  {title} {\enquote {\bibinfo {title} {Tropical cyclones
  and climate change},}\ }\href@noop {} {\bibfield  {journal} {\bibinfo
  {journal} {Nature Geoscience}\ }\textbf {\bibinfo {volume} {3}},\ \bibinfo
  {pages} {157--2010} (\bibinfo {year} {2010})}\BibitemShut {NoStop}%
\bibitem [{\citenamefont {Pall}\ \emph {et~al.}(2011)\citenamefont {Pall},
  \citenamefont {Aina}, \citenamefont {Stone}, \citenamefont {Stott},
  \citenamefont {Nozawa}, \citenamefont {Hilberts}, \citenamefont {Lohmann},\
  and\ \citenamefont {Allen}}]{pall2011anthropogenic}%
  \BibitemOpen
  \bibfield  {author} {\bibinfo {author} {\bibfnamefont {P.}~\bibnamefont
  {Pall}}, \bibinfo {author} {\bibfnamefont {T.}~\bibnamefont {Aina}}, \bibinfo
  {author} {\bibfnamefont {D.~A.}\ \bibnamefont {Stone}}, \bibinfo {author}
  {\bibfnamefont {P.~A.}\ \bibnamefont {Stott}}, \bibinfo {author}
  {\bibfnamefont {T.}~\bibnamefont {Nozawa}}, \bibinfo {author} {\bibfnamefont
  {A.~G.}\ \bibnamefont {Hilberts}}, \bibinfo {author} {\bibfnamefont
  {D.}~\bibnamefont {Lohmann}}, \ and\ \bibinfo {author} {\bibfnamefont
  {M.~R.}\ \bibnamefont {Allen}},\ }\bibfield  {title} {\enquote {\bibinfo
  {title} {Anthropogenic greenhouse gas contribution to flood risk in {E}ngland
  and {W}ales in autumn 2000},}\ }\href@noop {} {\bibfield  {journal} {\bibinfo
   {journal} {Nature}\ }\textbf {\bibinfo {volume} {470}},\ \bibinfo {pages}
  {382--385} (\bibinfo {year} {2011})}\BibitemShut {NoStop}%
\bibitem [{\citenamefont {Saha}\ \emph {et~al.}(2010)\citenamefont {Saha},
  \citenamefont {Moorthi}, \citenamefont {Pan}, \citenamefont {Wu},
  \citenamefont {Wang}, \citenamefont {Nadiga}, \citenamefont {Tripp},
  \citenamefont {Kistler}, \citenamefont {Woollen}, \citenamefont {Behringer}
  \emph {et~al.}}]{saha2010ncep}%
  \BibitemOpen
  \bibfield  {author} {\bibinfo {author} {\bibfnamefont {S.}~\bibnamefont
  {Saha}}, \bibinfo {author} {\bibfnamefont {S.}~\bibnamefont {Moorthi}},
  \bibinfo {author} {\bibfnamefont {H.-L.}\ \bibnamefont {Pan}}, \bibinfo
  {author} {\bibfnamefont {X.}~\bibnamefont {Wu}}, \bibinfo {author}
  {\bibfnamefont {J.}~\bibnamefont {Wang}}, \bibinfo {author} {\bibfnamefont
  {S.}~\bibnamefont {Nadiga}}, \bibinfo {author} {\bibfnamefont
  {P.}~\bibnamefont {Tripp}}, \bibinfo {author} {\bibfnamefont
  {R.}~\bibnamefont {Kistler}}, \bibinfo {author} {\bibfnamefont
  {J.}~\bibnamefont {Woollen}}, \bibinfo {author} {\bibfnamefont
  {D.}~\bibnamefont {Behringer}},  \emph {et~al.},\ }\bibfield  {title}
  {\enquote {\bibinfo {title} {The {NCEP} climate forecast system
  reanalysis},}\ }\href@noop {} {\bibfield  {journal} {\bibinfo  {journal}
  {Bulletin of the American Meteorological Society}\ }\textbf {\bibinfo
  {volume} {91}},\ \bibinfo {pages} {1015--1058} (\bibinfo {year}
  {2010})}\BibitemShut {NoStop}%
\bibitem [{\citenamefont {Maraun}\ \emph {et~al.}(2010)\citenamefont {Maraun},
  \citenamefont {Wetterhall}, \citenamefont {Ireson}, \citenamefont {Chandler},
  \citenamefont {Kendon}, \citenamefont {Widmann}, \citenamefont {Brienen},
  \citenamefont {Rust}, \citenamefont {Sauter}, \citenamefont {Theme{\ss}l}
  \emph {et~al.}}]{maraun2010precipitation}%
  \BibitemOpen
  \bibfield  {author} {\bibinfo {author} {\bibfnamefont {D.}~\bibnamefont
  {Maraun}}, \bibinfo {author} {\bibfnamefont {F.}~\bibnamefont {Wetterhall}},
  \bibinfo {author} {\bibfnamefont {A.}~\bibnamefont {Ireson}}, \bibinfo
  {author} {\bibfnamefont {R.}~\bibnamefont {Chandler}}, \bibinfo {author}
  {\bibfnamefont {E.}~\bibnamefont {Kendon}}, \bibinfo {author} {\bibfnamefont
  {M.}~\bibnamefont {Widmann}}, \bibinfo {author} {\bibfnamefont
  {S.}~\bibnamefont {Brienen}}, \bibinfo {author} {\bibfnamefont
  {H.}~\bibnamefont {Rust}}, \bibinfo {author} {\bibfnamefont {T.}~\bibnamefont
  {Sauter}}, \bibinfo {author} {\bibfnamefont {M.}~\bibnamefont {Theme{\ss}l}},
   \emph {et~al.},\ }\bibfield  {title} {\enquote {\bibinfo {title}
  {Precipitation downscaling under climate change: Recent developments to
  bridge the gap between dynamical models and the end user},}\ }\href@noop {}
  {\bibfield  {journal} {\bibinfo  {journal} {Reviews of Geophysics}\ }\textbf
  {\bibinfo {volume} {48}},\ \bibinfo {pages} {1--34} (\bibinfo {year}
  {2010})}\BibitemShut {NoStop}%
\bibitem [{\citenamefont {Emanuel}(1988)}]{emanuel1988toward}%
  \BibitemOpen
  \bibfield  {author} {\bibinfo {author} {\bibfnamefont {K.~A.}\ \bibnamefont
  {Emanuel}},\ }\bibfield  {title} {\enquote {\bibinfo {title} {Toward a
  general theory of hurricanes},}\ }\href@noop {} {\bibfield  {journal}
  {\bibinfo  {journal} {American Scientist}\ }\textbf {\bibinfo {volume}
  {76}},\ \bibinfo {pages} {370--379} (\bibinfo {year} {1988})}\BibitemShut
  {NoStop}%
\bibitem [{\citenamefont {Fritz}\ \emph {et~al.}(2009)\citenamefont {Fritz},
  \citenamefont {Blount}, \citenamefont {Thwin}, \citenamefont {Thu},\ and\
  \citenamefont {Chan}}]{fritz2009cyclone}%
  \BibitemOpen
  \bibfield  {author} {\bibinfo {author} {\bibfnamefont {H.~M.}\ \bibnamefont
  {Fritz}}, \bibinfo {author} {\bibfnamefont {C.~D.}\ \bibnamefont {Blount}},
  \bibinfo {author} {\bibfnamefont {S.}~\bibnamefont {Thwin}}, \bibinfo
  {author} {\bibfnamefont {M.~K.}\ \bibnamefont {Thu}}, \ and\ \bibinfo
  {author} {\bibfnamefont {N.}~\bibnamefont {Chan}},\ }\bibfield  {title}
  {\enquote {\bibinfo {title} {Cyclone {N}argis storm surge in {M}yanmar},}\
  }\href@noop {} {\bibfield  {journal} {\bibinfo  {journal} {Nature
  Geoscience}\ }\textbf {\bibinfo {volume} {2}},\ \bibinfo {pages} {448}
  (\bibinfo {year} {2009})}\BibitemShut {NoStop}%
\bibitem [{\citenamefont {Bakkensen}\ and\ \citenamefont
  {Mendelsohn}(2016)}]{bakkensen2016risk}%
  \BibitemOpen
  \bibfield  {author} {\bibinfo {author} {\bibfnamefont {L.~A.}\ \bibnamefont
  {Bakkensen}}\ and\ \bibinfo {author} {\bibfnamefont {R.~O.}\ \bibnamefont
  {Mendelsohn}},\ }\bibfield  {title} {\enquote {\bibinfo {title} {Risk and
  adaptation: evidence from global hurricane damages and fatalities},}\
  }\href@noop {} {\bibfield  {journal} {\bibinfo  {journal} {Journal of the
  Association of Environmental and Resource Economists}\ }\textbf {\bibinfo
  {volume} {3}},\ \bibinfo {pages} {555--587} (\bibinfo {year}
  {2016})}\BibitemShut {NoStop}%
\bibitem [{\citenamefont {Langewiesche}(2018)}]{langewiesche2018clock}%
  \BibitemOpen
  \bibfield  {author} {\bibinfo {author} {\bibfnamefont {W.}~\bibnamefont
  {Langewiesche}},\ }\bibfield  {title} {\enquote {\bibinfo {title} {`{T}he
  clock is ticking': inside the worse {U.S.} maritime disaster in decades},}\
  }\href@noop {} {\bibfield  {journal} {\bibinfo  {journal} {Vanity Fair}\ }
  (\bibinfo {year} {2018})}\BibitemShut {NoStop}%
\bibitem [{\citenamefont {Skamarock}\ \emph {et~al.}(2008)\citenamefont
  {Skamarock}, \citenamefont {Klemp}, \citenamefont {Dudhia}, \citenamefont
  {Gill}, \citenamefont {Barker}, \citenamefont {Duda}, \citenamefont {Huang},
  \citenamefont {Wang},\ and\ \citenamefont
  {Powers}}]{skamarock2008description}%
  \BibitemOpen
  \bibfield  {author} {\bibinfo {author} {\bibfnamefont {W.}~\bibnamefont
  {Skamarock}}, \bibinfo {author} {\bibfnamefont {J.}~\bibnamefont {Klemp}},
  \bibinfo {author} {\bibfnamefont {J.}~\bibnamefont {Dudhia}}, \bibinfo
  {author} {\bibfnamefont {D.}~\bibnamefont {Gill}}, \bibinfo {author}
  {\bibfnamefont {D.}~\bibnamefont {Barker}}, \bibinfo {author} {\bibfnamefont
  {M.}~\bibnamefont {Duda}}, \bibinfo {author} {\bibfnamefont {X.}~\bibnamefont
  {Huang}}, \bibinfo {author} {\bibfnamefont {W.}~\bibnamefont {Wang}}, \ and\
  \bibinfo {author} {\bibfnamefont {J.}~\bibnamefont {Powers}},\ }\bibfield
  {title} {\enquote {\bibinfo {title} {A description of the advanced research
  {WRF} version 3: {NCAR/TN-475}},}\ }\href@noop {} {\bibfield  {journal}
  {\bibinfo  {journal} {STR, NCAR technical note. NCAR, Boulder}\ } (\bibinfo
  {year} {2008})}\BibitemShut {NoStop}%
\bibitem [{\citenamefont {Judt}, \citenamefont {Chen},\ and\ \citenamefont
  {Berner}(2016)}]{judt2016predictability}%
  \BibitemOpen
  \bibfield  {author} {\bibinfo {author} {\bibfnamefont {F.}~\bibnamefont
  {Judt}}, \bibinfo {author} {\bibfnamefont {S.~S.}\ \bibnamefont {Chen}}, \
  and\ \bibinfo {author} {\bibfnamefont {J.}~\bibnamefont {Berner}},\
  }\bibfield  {title} {\enquote {\bibinfo {title} {Predictability of tropical
  cyclone intensity: scale-dependent forecast error growth in high-resolution
  stochastic kinetic-energy backscatter ensembles},}\ }\href@noop {} {\bibfield
   {journal} {\bibinfo  {journal} {Quarterly Journal of the Royal
  Meteorological Society}\ }\textbf {\bibinfo {volume} {142}},\ \bibinfo
  {pages} {43--57} (\bibinfo {year} {2016})}\BibitemShut {NoStop}%
\bibitem [{\citenamefont {Berner}\ \emph {et~al.}(2011)\citenamefont {Berner},
  \citenamefont {Ha}, \citenamefont {Hacker}, \citenamefont {Fournier},\ and\
  \citenamefont {Snyder}}]{berner2011model}%
  \BibitemOpen
  \bibfield  {author} {\bibinfo {author} {\bibfnamefont {J.}~\bibnamefont
  {Berner}}, \bibinfo {author} {\bibfnamefont {S.-Y.}\ \bibnamefont {Ha}},
  \bibinfo {author} {\bibfnamefont {J.}~\bibnamefont {Hacker}}, \bibinfo
  {author} {\bibfnamefont {A.}~\bibnamefont {Fournier}}, \ and\ \bibinfo
  {author} {\bibfnamefont {C.}~\bibnamefont {Snyder}},\ }\bibfield  {title}
  {\enquote {\bibinfo {title} {Model uncertainty in a mesoscale ensemble
  prediction system: Stochastic versus multiphysics representations},}\
  }\href@noop {} {\bibfield  {journal} {\bibinfo  {journal} {Monthly Weather
  Review}\ }\textbf {\bibinfo {volume} {139}},\ \bibinfo {pages} {1972--1995}
  (\bibinfo {year} {2011})}\BibitemShut {NoStop}%
\bibitem [{\citenamefont {Landsea}\ and\ \citenamefont
  {Franklin}(2013)}]{landsea2013atlantic}%
  \BibitemOpen
  \bibfield  {author} {\bibinfo {author} {\bibfnamefont {C.~W.}\ \bibnamefont
  {Landsea}}\ and\ \bibinfo {author} {\bibfnamefont {J.~L.}\ \bibnamefont
  {Franklin}},\ }\bibfield  {title} {\enquote {\bibinfo {title} {Atlantic
  hurricane database uncertainty and presentation of a new database format},}\
  }\href@noop {} {\bibfield  {journal} {\bibinfo  {journal} {Monthly Weather
  Review}\ }\textbf {\bibinfo {volume} {141}},\ \bibinfo {pages} {3576--3592}
  (\bibinfo {year} {2013})}\BibitemShut {NoStop}%
\bibitem [{\citenamefont {Chavas}, \citenamefont {Reed},\ and\ \citenamefont
  {Knaff}(2017)}]{chavas2017physical}%
  \BibitemOpen
  \bibfield  {author} {\bibinfo {author} {\bibfnamefont {D.~R.}\ \bibnamefont
  {Chavas}}, \bibinfo {author} {\bibfnamefont {K.~A.}\ \bibnamefont {Reed}}, \
  and\ \bibinfo {author} {\bibfnamefont {J.~A.}\ \bibnamefont {Knaff}},\
  }\bibfield  {title} {\enquote {\bibinfo {title} {Physical understanding of
  the tropical cyclone wind-pressure relationship},}\ }\href@noop {} {\bibfield
   {journal} {\bibinfo  {journal} {Nature Communications}\ }\textbf {\bibinfo
  {volume} {8}},\ \bibinfo {pages} {1--11} (\bibinfo {year}
  {2017})}\BibitemShut {NoStop}%
\bibitem [{\citenamefont {Zhai}\ and\ \citenamefont
  {Jiang}(2014)}]{zhai2014dependence}%
  \BibitemOpen
  \bibfield  {author} {\bibinfo {author} {\bibfnamefont {A.~R.}\ \bibnamefont
  {Zhai}}\ and\ \bibinfo {author} {\bibfnamefont {J.~H.}\ \bibnamefont
  {Jiang}},\ }\bibfield  {title} {\enquote {\bibinfo {title} {Dependence of
  {US} hurricane economic loss on maximum wind speed and storm size},}\
  }\href@noop {} {\bibfield  {journal} {\bibinfo  {journal} {Environmental
  Research Letters}\ }\textbf {\bibinfo {volume} {9}},\ \bibinfo {pages} {1--9}
  (\bibinfo {year} {2014})}\BibitemShut {NoStop}%
\bibitem [{\citenamefont {of~the Federal Coordinator~for
  Meteorological~Services}\ and\ \citenamefont
  {Research}(2010)}]{ofcm2010national}%
  \BibitemOpen
  \bibfield  {author} {\bibinfo {author} {\bibfnamefont {O.}~\bibnamefont
  {of~the Federal Coordinator~for Meteorological~Services}}\ and\ \bibinfo
  {author} {\bibfnamefont {S.}~\bibnamefont {Research}},\ }\href@noop {}
  {\enquote {\bibinfo {title} {National hurricane operations plan:
  {FCM-P12-2012}},}\ }\bibinfo {type} {Tech. Rep.}\ (\bibinfo  {institution}
  {National Oceanic and Atmospheric Administration},\ \bibinfo {year}
  {2010})\BibitemShut {NoStop}%
\bibitem [{\citenamefont {Lee}\ \emph {et~al.}(2016)\citenamefont {Lee},
  \citenamefont {Tippett}, \citenamefont {Sobel},\ and\ \citenamefont
  {Camargo}}]{lee2016rapid}%
  \BibitemOpen
  \bibfield  {author} {\bibinfo {author} {\bibfnamefont {C.-Y.}\ \bibnamefont
  {Lee}}, \bibinfo {author} {\bibfnamefont {M.~K.}\ \bibnamefont {Tippett}},
  \bibinfo {author} {\bibfnamefont {A.~H.}\ \bibnamefont {Sobel}}, \ and\
  \bibinfo {author} {\bibfnamefont {S.~J.}\ \bibnamefont {Camargo}},\
  }\bibfield  {title} {\enquote {\bibinfo {title} {Rapid intensification and
  the bimodal distribution of tropical cyclone intensity},}\ }\href@noop {}
  {\bibfield  {journal} {\bibinfo  {journal} {Nature Communications}\ }\textbf
  {\bibinfo {volume} {7}},\ \bibinfo {pages} {1--5} (\bibinfo {year}
  {2016})}\BibitemShut {NoStop}%
\bibitem [{\citenamefont {Montgomery}\ and\ \citenamefont
  {Smith}(2014)}]{montgomery2014paradigms}%
  \BibitemOpen
  \bibfield  {author} {\bibinfo {author} {\bibfnamefont {M.~T.}\ \bibnamefont
  {Montgomery}}\ and\ \bibinfo {author} {\bibfnamefont {R.~K.}\ \bibnamefont
  {Smith}},\ }\bibfield  {title} {\enquote {\bibinfo {title} {Paradigms for
  tropical cyclone intensification},}\ }\href@noop {} {\bibfield  {journal}
  {\bibinfo  {journal} {Australian Meteorological and Oceanographic Journal}\
  }\textbf {\bibinfo {volume} {64}},\ \bibinfo {pages} {37--66} (\bibinfo
  {year} {2014})}\BibitemShut {NoStop}%
\bibitem [{\citenamefont {Berner}\ \emph {et~al.}(2017)\citenamefont {Berner},
  \citenamefont {Achatz}, \citenamefont {Batt{\'e}}, \citenamefont {Bengtsson},
  \citenamefont {C{\'a}mara}, \citenamefont {Christensen}, \citenamefont
  {Colangeli}, \citenamefont {Coleman}, \citenamefont {Crommelin},
  \citenamefont {Dolaptchiev} \emph {et~al.}}]{berner2017stochastic}%
  \BibitemOpen
  \bibfield  {author} {\bibinfo {author} {\bibfnamefont {J.}~\bibnamefont
  {Berner}}, \bibinfo {author} {\bibfnamefont {U.}~\bibnamefont {Achatz}},
  \bibinfo {author} {\bibfnamefont {L.}~\bibnamefont {Batt{\'e}}}, \bibinfo
  {author} {\bibfnamefont {L.}~\bibnamefont {Bengtsson}}, \bibinfo {author}
  {\bibfnamefont {A.~d.~l.}\ \bibnamefont {C{\'a}mara}}, \bibinfo {author}
  {\bibfnamefont {H.~M.}\ \bibnamefont {Christensen}}, \bibinfo {author}
  {\bibfnamefont {M.}~\bibnamefont {Colangeli}}, \bibinfo {author}
  {\bibfnamefont {D.~R.}\ \bibnamefont {Coleman}}, \bibinfo {author}
  {\bibfnamefont {D.}~\bibnamefont {Crommelin}}, \bibinfo {author}
  {\bibfnamefont {S.~I.}\ \bibnamefont {Dolaptchiev}},  \emph {et~al.},\
  }\bibfield  {title} {\enquote {\bibinfo {title} {Stochastic parameterization:
  Toward a new view of weather and climate models},}\ }\href@noop {} {\bibfield
   {journal} {\bibinfo  {journal} {Bulletin of the American Meteorological
  Society}\ }\textbf {\bibinfo {volume} {98}},\ \bibinfo {pages} {565--588}
  (\bibinfo {year} {2017})}\BibitemShut {NoStop}%
\bibitem [{\citenamefont {Emanuel}, \citenamefont {Sundararajan},\ and\
  \citenamefont {Williams}(2008)}]{emanuel2008hurricanes}%
  \BibitemOpen
  \bibfield  {author} {\bibinfo {author} {\bibfnamefont {K.}~\bibnamefont
  {Emanuel}}, \bibinfo {author} {\bibfnamefont {R.}~\bibnamefont
  {Sundararajan}}, \ and\ \bibinfo {author} {\bibfnamefont {J.}~\bibnamefont
  {Williams}},\ }\bibfield  {title} {\enquote {\bibinfo {title} {Hurricanes and
  global warming: Results from downscaling {IPCC AR4} simulations},}\
  }\href@noop {} {\bibfield  {journal} {\bibinfo  {journal} {Bulletin of the
  American Meteorological Society}\ }\textbf {\bibinfo {volume} {89}},\
  \bibinfo {pages} {347--368} (\bibinfo {year} {2008})}\BibitemShut {NoStop}%
\end{thebibliography}%

\end{document}